\documentclass[11pt]{article}
\usepackage{placeins}
\usepackage{amsmath}
\usepackage{amsthm}
\usepackage{amssymb}
\usepackage{bigints}
\usepackage{graphicx}
\usepackage{keyval}
\usepackage{booktabs}
\usepackage{float}
\usepackage{setspace}
\usepackage{mathrsfs}
\usepackage{subfigure}
\usepackage{longtable}
\usepackage{verbatim}
\usepackage{arydshln}
\usepackage{enumerate}
\usepackage[usenames, dvipsnames]{color}
\usepackage{color}
\usepackage{footmisc}
\usepackage[authoryear]{natbib}
\usepackage{tikz}
\usepackage{soul}
\usepackage{multirow}
\usepackage{framed} 
\usepackage{color, colortbl}
\definecolor{LightCyan}{rgb}{0.88,1,1}
\definecolor{LightOrange}{rgb}{1,.84,.5}
\definecolor{LightGray}{gray}{0.9}
\usepackage{tablefootnote}
\usepackage{enumitem}
\usepackage{xcolor}
\usepackage{float}
\usepackage{pdflscape}
\DeclareMathOperator{\sign}{sign}
\usepackage{pgfplots}
\pgfplotsset{compat=1.15,
             height=0.8\columnwidth,
             width=\columnwidth
             }\usepgfplotslibrary{ternary}

\usepackage{diagbox}
\usepackage{slashbox}
\usepackage{multirow}
\usepackage{hyperref}
\hypersetup{
	pdfstartview = FitH,
	pdfauthor = {...},
	pdftitle = {...},
	pdfkeywords = {...; ...; ...; ...},
	colorlinks = true,
	linkcolor = orange,
	urlcolor = black,
	citecolor = blue,
	linktocpage=true}
	
\allowdisplaybreaks

\usepackage{geometry}
\graphicspath{{Plots/}}
\numberwithin{equation}{section}
\newtheorem{proposition}{Proposition}
\newtheorem*{proposition*}{Proposition}

\newtheorem{lemma}{Lemma}

\numberwithin{assumption}{section}

\newtheorem{definition}{Definition}

\theoremstyle{definition}

\DeclareMathOperator*{\argmax}{arg\,max}

\begin{document}
\begin{titlepage}
\pagenumbering{gobble}
\title{\LARGE \textbf{Distributive Politics, Representation, and Redistricting}\thanks{All errors are our own.}}
\author{\textbf{Thomas Groll}\thanks{School of International and Public Affairs, Columbia University, New York, NY; tgroll@columbia.edu.} \\ Columbia University \\ \and \textbf{Sharyn O'Halloran}\thanks{School of International and Public Affairs and Political Science, Columbia University, New York, NY, USA; Department of Economics and Political Science, Trinity College Dublin, Dublin, Ireland; so33@columbia.edu.} \\ Columbia University \\ Trinity College Dublin}

\date{March 31, 2026}
\maketitle

\begin{abstract}
We develop a theory of distributive competition under redistricting that explains both electoral outcomes and the equilibrium allocation of policy benefits by endogenizing voter pivotality. In a multi-district model with primaries, general elections, and group-targeted transfers, districting shapes political influence through two channels: a selection channel for descriptive representation (who wins office) and a competition channel for substantive representation (who receives policy benefits). District composition alters candidate matchups, shifting voter responsiveness and political leverage, and each channel alone yields distinct predictions about whether packing or cracking voters is optimal. For minority voters, the welfare effects of districting depend on electoral leverage, preferences over descriptive versus partisan representation, primary rules, and competitiveness. The channels align on packing when minorities are electorally weak and value descriptive representation, and align on cracking when minorities are electorally pivotal and prioritize partisan outcomes. When the channels diverge, or when endogenous feedback reshapes electoral leverage, minority welfare can be nonmonotonic in voter concentration. Our results identify when majority–minority districts enhance minority welfare and when dispersion strengthens political influence.
\end{abstract}

\textit{Keywords:} Distributive Politics, Redistricting, Electoral Competition, Political Influence, Minority Representation, Voting Rights Act \\


\end{titlepage}
	\setlength{\parindent}{.25in}

\doublespacing

\newpage
\pagenumbering{arabic}	
	\setcounter{page}{1}		

\section{Introduction}\label{section_intro}

Electoral institutions shape both who wins office and how policy benefits are distributed across voters. Majority–minority districts are often justified as a means of enhancing minority representation. Yet concentrating minority voters in a few districts may increase their probability of electing preferred candidates while reducing their leverage over policy benefits in surrounding districts. Whether packing minority voters enhances or weakens their overall welfare, therefore, depends not only on who wins office but also on how district composition shapes electoral competition and a group's leverage over the allocation of policy benefits. Existing models of redistricting focus primarily on electoral outcomes, while models of distributive politics analyze the allocation of policy benefits independently of district maps. 

We connect these literatures by showing how district composition simultaneously shapes candidate selection, voter responsiveness, and the equilibrium distribution of policy benefits. We develop a model in which district composition affects both electoral outcomes and the distribution of policy benefits by endogenizing voter responsiveness and group political leverage. In a multi-district extension of \citet{dixitlondregan1996} with primaries, general elections, and group-targeted transfers, district composition affects both candidate matchups and voter responsiveness. Primaries matter because district composition influences both vote shares and which candidates compete in the general election, and these matchups determine group leverage in distributive competition. 

Modeling voter groups and primaries also reflects the Voting Rights Act's institutional logic, which evaluates whether identifiable minority constituencies can elect preferred candidates and exercise meaningful influence over policy outcomes. By altering which candidates face each other, districting shifts marginal voter responsiveness and a group's political leverage in distributive competition. This structure distinguishes between a \emph{selection} channel (how district composition affects which candidates reach and win the general election) and a \emph{competition} channel (how it shapes policy targeting incentives), allowing their effects to interact in general equilibrium. Although our framework applies broadly, we focus on minority representation, where the tension between seat share and policy leverage is especially salient. More generally, the mechanism applies whenever institutional rules alter candidate competition and voter responsiveness, thereby redistributing political influence across groups.

The results proceed in four steps. We first show how endogenous candidate matchups affect distributive outcomes in equilibrium. We then characterize the competition and selection channels before examining how endogenous feedback arises as matchups adjust, and how these general equilibrium interactions highlight sources of nonmonotonicity in minority welfare. First, in the competition channel, which focuses on policy targeting under fixed matchups, the optimal direction of minority concentration depends on group leverage in distributive competition. When minority voters have low leverage, concentration strengthens their influence. When they are electorally pivotal, dispersion spreads minority influence across districts. Second, in the selection channel, which focuses on candidate choice, the welfare effects of packing or cracking depend on primary rules, preferences for descriptive versus partisan representation, and whether minority candidates improve general-election success. When minority voters prioritize electing minority candidates, concentration increases welfare. When their objective is to prevent partisan opponents from winning office, dispersion may instead be beneficial.

Third, when candidate matchups adjust endogenously, district composition feeds back into voter responsiveness and group leverage, reinforcing, dampening, or overturning the channels’ benchmark predictions. In electorally safe districts, compositional changes have limited effects on matchups and responsiveness; the interaction is second-order, and alignment is preserved. In competitive districts, small compositional changes can shift matchups and marginal responsiveness, making group leverage first-order and potentially reversing the fixed-matchup prediction. Fourth, in general equilibrium, the interaction between the selection and competition channels implies that optimal districting depends on both benchmark effects and endogenous feedback. As a result, minority welfare can be nonmonotonic in voter concentration when the channels diverge or when electoral responsiveness overturns benchmark effects.

Taken together, our analysis with endogenous electoral power shows that redistricting is not merely an allocation of voters across districts but an institutional design problem that shapes the structure of political competition itself. We show how the welfare consequences of packing and cracking depend on minority power (their electoral leverage) and preference weights (descriptive versus partisan) as well as on the primary system and the candidates running for office. These welfare implications are especially severe in competitive districts where alignments or divergence between descriptive and substantive representation can be amplified, dampened, or even reversed.

Taken together, these results show that districting shapes not only electoral outcomes but also the equilibrium distribution of policy benefits by altering voter responsiveness and the political leverage of voter groups. The model's findings have direct policy implications: reforms that prioritize minority seat share, for example, by creating majority-minority districts, may inadvertently reduce distributive leverage, whereas dispersion can amplify policy competition under certain conditions. Our framework also yields comparative statics linking district composition, electoral competitiveness, and minority leverage, generating testable predictions about when packing or dispersion increases minority political influence.

The remainder of the paper proceeds as follows. Section \ref{section_intro-literature} situates our analysis within the literature on redistricting and representation. Sections \ref{section_model} and \ref{subsection_equilibrium-platforms} present the general model and characterize equilibrium policy competition. Section \ref{section_substantive-representation} analyzes the isolated competition channel and illustrates substantive representation. Section \ref{section_descriptive-representation} analyzes the isolated selection channel and illustrates descriptive representation. Section \ref{section_tradeoffs-representation} analyzes the general equilibrium effects with interactions between the two channels, highlighting trade-offs. Section \ref{section_conclusion} discusses our results and concludes with policy implications.

\subsection{Related Literature}\label{section_intro-literature}

Our analysis connects to three literatures: distributive politics, redistricting, and minority representation. These strands are often studied in isolation, focusing either on how electoral competition allocates policy benefits, how parties maximize vote shares and seats, or the trade-offs facing racial and ethnic minorities. 

The distributive politics literature studies how electoral competition shapes policy targeting across voter groups \citep{lindbeckweibull1987, myerson1993, dixitlondregan1995, dixitlondregan1996}. These models emphasize how marginal voter responsiveness determines the allocation of policy benefits. We extend this framework to a multi-district environment in which district composition reshapes voter responsiveness and the electoral leverage of voter groups. The group-targeting focus also reflects the institutional logic of the Voting Rights Act, which evaluates political representation in terms of identifiable voter groups, especially protected racial minorities, and their ability to exercise collective political influence. By linking distributive competition to district composition, our framework shows how electoral institutions affect not only who wins office but also the equilibrium distribution of policy benefits.

While models of distributive politics explain how electoral competition determines policy targeting, the redistricting literature focuses primarily on how gerrymanders manipulate boundaries to create seat–vote biases \citep{tufte1973, king1989, gelmanking1990, lublinhandleybrunellgrofman2020}, protect incumbents \citep{coxkatz2002, ansolabeheresnyder2004}, and consolidate party power \citep{butlercain1991, Issacharoff_2002, persily2002}. In contrast, the racial redistricting literature investigates Black officeholding in the South \citep{davidsongrofman1994}, the tension between descriptive and substantive representation \citep{cameronohalloran1996, epsteinohalloran1999AJPS, lublin1999book, EpsteinHerronOHalloran_2007}, partisan shifts in Congress \citep{lublinvoss2000}, and the broader consequences for polarization \citep{stephanopoulos2016}.\footnote{\citet{canon2022} and \citet{bertall2024} analyze the legal evolution of descriptive and substantive representation.} 

Empirical research highlights persistent trade-offs in minority redistricting. \citet{jeongshenoy2022} document the ``packing-and-cracking'' of minority voters, showing how Republican redistricting concentrates minority voters in majority–minority districts. While \citet{Canon_1999} emphasizes the descriptive benefits of such districts, \citet{cameronohalloran1996} and \citet{lublin1997apr} argue that concentration may dilute substantive policy influence.\footnote{\citet{Issacharoff_2002, GrofmanHandleyNiemi1991, pildesetal_2006} caution that such influence is difficult to measure, suggesting that Section 5 of the Voting Rights Act may have increased minority representation while weakening broader policy leverage; see also \citet{epsteinohalloran1999AJPS}.} The magnitude of this trade-off depends on crossover voting -- i.e., nonminority voters' willingness to support minority candidates. Here, the evidence is mixed: \citet{ansolabeherepersilystewart2010} find such support prevalent, whereas \citet{lublinhandleybrunellgrofman2020} suggest it is declining. The literature's persistent ambiguity underscores the need for a theoretical framework to evaluate optimal strategies across varying levels of voter polarization.

Formal models of redistricting largely analyze how strategic gerrymanders allocate voters across districts to shape electoral outcomes. Early work focused on seat maximization \citep{musgrove1977, owengrofman1988}, while subsequent studies examine how majority–minority mandates constrain conservative gerrymanders \citep{shotts2001}, how cracking becomes inefficient under preference uncertainty \citep{friedmanholden2008}, and how two-party competition generates partisan voter segregation \citep{gulpesendorfer2010}. Recent work in economics introduces additional mechanisms, including turnout differentials \citep{boutonetalnber2023}, preference uncertainty \citep{kolotilinwolitzky2026}, and endogenous candidate selection \citep{moscariello2025}. In a related normative strand, \citet{coateknight2007} analyze socially optimal districting from a social planner's perspective who balances representation and policy outcomes, and \citet{bracco2013} extends their analysis by letting parties choose platforms after districting. 

We depart from this tradition by integrating distributive competition with districting, allowing district composition to reshape voter responsiveness and the political leverage of voter groups across electoral stages. As a result, districting affects not only electoral outcomes but also the equilibrium distribution of policy benefits across voter groups. A related framework studies distributive competition under fixed group power \citep{grollohalloran2024}. By contrast, we allow redistricting to endogenously determine group power, generating feedback between districting and distributive competition.

Our theoretical tensions are reflected in the jurisprudence of the Voting Rights Act of 1965, which initially prioritized descriptive representation through majority–minority districts. Subsequent jurisprudence, however, has grappled with the ``unpacking'' logic, recognizing that concentration may dilute substantive influence.\footnote{The tension between descriptive and substantive representation has been central to the evolution of redistricting law. In \textit{Georgia v.~Ashcroft} (2003), the Court accepted the logic of “unpacking,” reasoning that spreading minority voters could expand overall policy influence even at the expense of electing fewer minority representatives. In \textit{Cooper v.~Harris} (2017), the Court struck down racial “packing,” emphasizing that concentration does not always enhance minority power, particularly when white crossover voting reduces polarization. Finally, in \textit{Rucho v.~Common Cause} (2019), the Court held that partisan gerrymandering was nonjusticiable, effectively separating racial from partisan claims in law even as they remain intertwined in practice. Empirically, \citet{gimpeletal2021} show that redistricting litigation has increasingly shifted from federal to state courts, concentrated in regions with large minority populations, underscoring how geography and partisanship structure these trade-offs.} Our analysis provides a formal benchmark to evaluate these legal and normative debates, clarifying when concentration enhances representation and when it weakens policy influence.

Although we focus on minority representation, the model applies more broadly to institutional design problems in electoral competition where institutions shape voter responsiveness and group leverage. Existing redistricting models largely focus on electoral outcomes (who wins office), whereas our approach also analyzes how districting shapes the allocation of policy benefits. Our framework integrates these literatures by showing how districting affects both electoral outcomes and the equilibrium distribution of policy benefits. It provides a method for evaluating when packing or dispersing minority voters enhances political influence.

\section{Model}\label{section_model}

We adapt the electoral competition framework of \citet{dixitlondregan1996} to a multi-district setting with endogenous candidate selection, endogenous voter group power, and partisan and racial redistricting. Voters have ideological attachments to candidates, and candidates compete by promising group-specific policy benefits. Voters have heterogeneous preferences over candidates and policy benefits, and, as a group, may favor candidates differently. The structure captures the trade-off between the ideological utility of candidate identity (\emph{descriptive} and \emph{partisan representation}) and the material utility of targeted transfers (\emph{substantive representation}). A designer takes this trade-off into account when proposing districting schemes and alters welfare through two channels: (i) the \emph{selection channel}, which determines the identity of elected officials, and (ii) the \emph{competition channel}, which alters the marginal incentives for candidates to target specific groups.

\subsection{Districts}\label{subsection_districts}

Consider a population of $N$ voters partitioned into identifiable groups $\Theta$. We focus on a state divided along ethnic and partisan lines, with voter types $i \in \Theta = \{mD,nD,R\}$ denoting minority Democrats, nonminority Democrats, and Republicans. Statewide group populations are $N_i$ with $\sum_i N_i=N$. The state is divided into $K$ districts denoted by $k=1,...,K$.

Each district $k$ consists of voter group counts with a vector $\mathbf{N_k}=(N_{mD,k},N_{nD,k},N_{R,k})$ with $\sum_{i}N_{i,k}=N_k$ and corresponding group shares $s_{i,k}=N_{i,k}/N_k$. A valid districting scheme satisfies (i) equal district populations, $N_k=N/K$ $\forall k$, and (ii) group totals across districts match the statewide population, $\sum_k N_{i,k}=N_i$. A districting scheme is a vector with $\mathbf{D}=(\mathbf{N_1},...,\mathbf{N_K})$. 

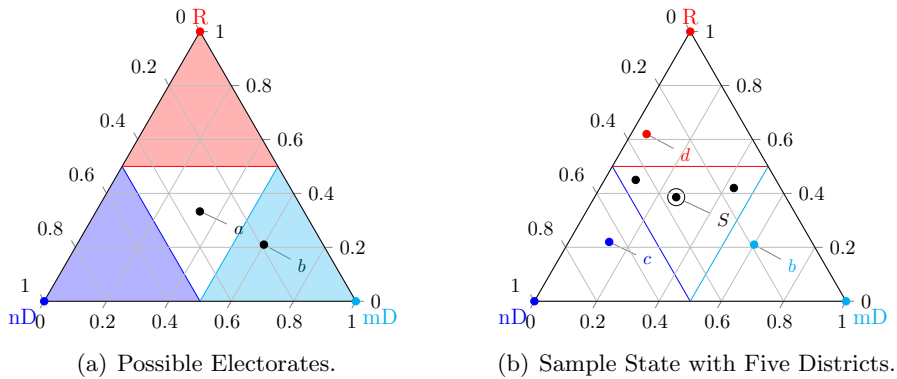
\begin{figure}[th]
    \centering
    \subfigure[Possible Electorates.]{\label{fig:IntroTriangle}\scalebox{0.7}{\begin{tikzpicture}
\begin{ternaryaxis}[
ternary limits relative=false,
xlabel=\large R,
xlabel style={
at={(axis cs:1,0,0)},
anchor=south,color=red},
ylabel=\large nD,
ylabel style={
at={(axis cs:0,1,0)},
anchor=north east,color=blue}, 
zlabel=\large mD,
zlabel style={
at={(axis cs:0,0,1)},
anchor=north west,color=cyan},
width=7.5cm,
height=7.5cm,
xmin=0,
xmax=1,
ymin=0,
ymax=1,
zmin=0,
zmax=1,
clip=false,
disabledatascaling,
area style,
]
\addplot3[only marks, mark options={red}]
table {
1    0    0
};
\addplot3[only marks, mark options={cyan}, ]
table {
0    0    1
};
\addplot3[only marks, mark options={blue}]
table {
0    1    0
};
\addplot3 [red] coordinates {
(0.5, 0.5, 0)
(0.5, 0, 0.5)
};
\addplot3 [blue] coordinates {
(0.5, .5, 0)
(0, 0.5, 0.5)
};
\addplot3 [cyan] coordinates {
(0.5, 0, 0.5)
(0, 0.5, 0.5)
};
\addplot3[only marks] table {
.333 0.333 0.333
.21    .19    .6
};
\node[inner sep=0.5pt,circle,draw,fill=white,
	  pin=-15:\small $a$] at (axis cs:0.3333,0.3333,0.3333) {};
\node[inner sep=0.5pt,circle,draw,fill=white,
	  pin=-15:\small $b$] at (axis cs:0.21,0.19,0.6) {};
		\fill[cyan,opacity=0.3] (axis cs: 0,0.5) -- (axis cs:0.5,0.0) -- (axis cs:0,0) -- cycle;
		\fill[blue,opacity=0.3] (axis cs: 0,1) -- (axis cs:0.5,0.5) -- (axis cs:0,0.5) -- cycle;
		\fill[red,opacity=0.3] (axis cs: 1,0) -- (axis cs:0.5,0.5) -- (axis cs:0.5,0) -- cycle;
\end{ternaryaxis}
\end{tikzpicture}}}
    \hspace{0.05\textwidth}
    \subfigure[Sample State with Five Districts.]{\label{fig:SampleState}\scalebox{0.7}{\begin{tikzpicture}
\begin{ternaryaxis}[
ternary limits relative=false,
xlabel=\large R,
xlabel style={
at={(axis cs:1,0,0)},
anchor=south,color=red},
ylabel=\large nD,
ylabel style={
at={(axis cs:0,1,0)},
anchor=north east,color=blue}, 
zlabel=\large mD,
zlabel style={
at={(axis cs:0,0,1)},
anchor=north west,color=cyan},
width=7.5cm,
height=7.5cm,
xmin=0,
xmax=1,
ymin=0,
ymax=1,
zmin=0,
zmax=1,
clip=false,
disabledatascaling,
area style,
]
\addplot3[only marks, mark options={red}]
table {
1    0    0
0.62  .33  .05
};
\addplot3[only marks, mark options={cyan}, ]
table {
0    0    1
.21  .19  .6
};
\addplot3[only marks, mark options={blue}]
table {
0    1    0
.22  .65  0.13
};
\addplot3 [red] coordinates {
(0.5, 0.5, 0)
(0.5, 0, 0.5)
};
\addplot3 [blue] coordinates {
(0.5, .5, 0)
(0, 0.5, 0.5)
};
\addplot3 [cyan] coordinates {
(0.5, 0, 0.5)
(0, 0.5, 0.5)
};
\addplot3[only marks] table {
0.386  0.352  0.262
.45  0.45  0.1
.42  0.14  0.43
};
\node[inner sep=1.25pt,circle=black,draw=black,fill=white,
	  pin=-15:\small $S$] at (axis cs:0.386, 0.352, 0.262) { $\cdot$};
\node[inner sep=0.5pt,circle,draw,text=red,
	  pin=-15:\small \textcolor{cyan}{$b$}] at (axis cs:0.21,0.19,0.6) {};
\node[inner sep=0.5pt,circle,draw,text=red,
	  pin=-15:\small \textcolor{blue}{$c$}] at (axis cs:0.22,0.65,0.13) {};
\node[inner sep=0.5pt,circle,draw,text=red,
	  pin=-15:\small \textcolor{red}{$d$}] at (axis cs:0.62,0.33,0.05) {};						
\end{ternaryaxis}
\end{tikzpicture}}}
    \caption{Possible Electorates and Districts.}
    \label{fig:triangles-districts}
\end{figure}

The composition of any district can be represented as a point in the two-dimensional simplex $S^2$ as illustrated Figure~\ref{fig:triangles-districts} using population shares in the district, $s_{i,k}$. The vertices correspond to homogeneous electorates, while the interior represents heterogeneous districts. Figure~\ref{fig:IntroTriangle} illustrates the political geography: district $b$ denotes a majority-minority district, while district $a$ illustrates no single-group majority. Figure~\ref{fig:SampleState} depicts a hypothetical five-district plan for a state with statewide populations $S=(s_{mD},s_{nD},s_{R})$.\footnote{See \ref{section_appendix_numerics_state} for numerical details.}

\subsection{Candidates and Elections}\label{subsection_candidates-elections}

In each district, candidates of types $j\in\Theta$ compete for a single seat. Candidates maximize their expected vote share by choosing a platform $T_{i,j,k}$, representing candidate $j$'s share of legislative benefits promised to group $i$ in district $k$. Platforms must satisfy the budget constraint $\sum_{i}T_{i,j,k}=1$. Redistricting affects outcomes by altering district group sizes $N_{i,k}$, thereby changing electoral incentives via the competition channel.

Elections proceed in two stages. First, a Democratic primary pits an $mD$ candidate against an $nD$ candidate. Second, the primary winner faces a Republican in the general election. This two-stage structure forces Democratic candidates to clear a selection threshold within their own party before competing for voters in the general election. To obtain smooth mappings from vote shares to electoral outcomes, we adopt probabilistic voting. If $v$ denotes a candidate's expected vote share, the probability of winning an election is given by a linear function $\Psi(v)=v$, with $\Psi(0)=0$, $\Psi(1)=1$, and $\Psi(1-v)=1-\Psi(v)$. We use $\Psi_j^e(k)$ to denote the probability that candidate $j$ wins election stage $e$ in district $k$, and $\Psi_j(k)$ to denote the probability that candidate $j$ wins district $k$ overall.\footnote{Note that expected vote shares aggregate smoothly across district group sizes and that $\Psi(v)=v$. Hence, the winning probabilities are smooth functions on the simplex $S^2$, with group-specific turnout or registration entering only as reduced-form scalars.} State primaries can be either closed or open.\footnote{We write $\Psi_{j}(k)$ for a state with closed primaries ($i'\in\{mD,nD\}$) and $\tilde{\Psi}_{j}(k)$ for a state with open primaries ($i\in\{mD,nD,R\}$).}

\subsection{Legislative Policies}\label{subsection_leg-policies}

After elections, the legislature allocates a fixed budget of $K$ dollars across districts. If district $k$ elects a representative of type $j$ with platform $T_{j,k}$, group $i$ in that district receives total benefits $T_{i,j,k} B_k$. Assuming a standard \citet{baronferejohn1989} bargaining process where recognition probabilities are proportional to seat shares, each district captures an equal expected share of the state budget -- i.e., $B_k=1$. The per-capita transfer to a voter in group $i$  is $b_{i,j,k} = T_{i,j,k}/N_{i,k}$ with $T_{i,j,k}=b_{i,j,k}=0$ for $N_{i,k}=0$.

\subsection{Voters}\label{subsection_voters}

Following \cite{dixitlondregan1996}, voters belong to an identifiable group $i\in\Theta$ and share common preferences over consumption, but have heterogeneous ideological affinities to candidates. We allow for candidate-specific group ideological payoffs, so that the relevant ideological term is election- and matchup-specific. Voters derive utility from two sources: (i) descriptive representation, arising from the identity or type of the elected candidate, and (ii) substantive representation, arising from policy benefits for consumption through targeted transfers.

The utility of a voter in group $i$ in district $k$, when candidate $j$ offers per-capita benefits $b_{i,j,k}$, is
\begin{equation}\label{voter-utility}
U_i(j,k)=\mu_{i,j}+u_i(b_{i,j,k}),
\end{equation} 
where $u_i(b)=\kappa_i\frac{b^{1-\epsilon}}{1-\epsilon}$, $ \mu_{i,j}$ is a voter group $i$'s mean ideological affinity for candidate $j$, $\kappa_i$ weighs the relative importance of consumption to ideology, and $\epsilon$ determines the curvature of consumption utility (degree of diminishing returns to consumption) with $\epsilon > 0$ and $\epsilon \neq 1$:\footnote{As in \cite{dixitlondregan1996}, the isoelastic utility function is adopted for tractability and closed-form solutions. As shown in Appendix~\ref{appendix_proof-lemma-platforms}, the qualitative targeting mechanism extends to any increasing, strictly concave consumption utility, and we only employ the constant relative risk aversion specifications for group power expressions and comparative statics. We also assume throughout that voter groups may differ in their consumption and ideological preferences, but neglect within-group differences across districts.}
\begin{equation}\label{eq:utility_consumption_FOC-SOC}
u_i'(b) = \kappa_i b^{-\epsilon} > 0 \text{ and } u_i''(b) = -\epsilon \kappa_i b^{-\epsilon-1} <0.
\end{equation}

Within each group $i$, voters differ in their ideological affinity between the two candidates in any election. Specifically, we assume 
\begin{equation}
\theta \thicksim \Phi_i(.),
\end{equation}
where individual voters differ in $\theta$. 

\paragraph{Voting} In an election of type $e$ in district $k$ between two candidates, denoted 1 and 2, voters are assumed to cast their ballots sincerely for the candidate who offers them the highest utility. A voter in group $i$ with a preference $\theta$ for Candidate 2 votes for Candidate 1 iff 
\begin{equation}
U_{i} (1,k) - U_{i} (2,k) > \theta \text{ }\Leftrightarrow \text{ } \theta < x_i^e(k),
\end{equation}
where 
\begin{equation}
x_i^e(k) = \mu_i^e + \kappa_i \frac{b_{i,1,k}^{1-\epsilon}-b_{i,2,k}^{1-\epsilon}}{1-\epsilon} \text{ with } \mu_i^e \equiv \mu_{i,1} -\mu_{i,2}.
\end{equation}
Because ideological payoffs are candidate- and matchup-specific in our framework, this cut-off is election-specific and varies for group $i$ in district $k$ across election stages and candidate identities. Hence, the first term $\mu_i^e$ illustrates descriptive representation, and the second term illustrates substantive representation.\footnote{At each electoral stage, only differences $\mu_{i,1} -\mu_{i,2}$ matter, and we denote this election-specific ideological bias by $\mu_i^e$ in the equilibrium analysis.} Candidates influence voter behavior through this threshold only through the promised transfers, $b_{i,j,k}$, which affect consumption and the group's marginal utility. A marginal increase in promised transfers shifts the vote share in proportion to the group's marginal utility $u_i'(b)$. Hence, groups with high marginal utility or low ideological commitment are ``cheaper" to buy; groups that are far from the indifference point in that election (larger $|\mu_i^e|$) are expensive to move, though they might be less expensive to move in a different election due to a different candidate matchup.

The above specification highlights the central trade-off. The fixed component $\mu_i^e$ determines the \emph{baseline probability} of a group supporting a candidate (descriptive and partisan representation). The marginal utility $u_i'(b) = \kappa_i b_{i,j,k}^{-\epsilon}$ determines the \emph{responsiveness} of that support to additional promises (substantive representation).

\paragraph{Votes} Given $N_{i,k}$ voters of type $i$, Candidate 1 will receive $N_{i,k}\Phi_i(x_i^e(k))$ votes from group $i$, with total votes of:
\begin{equation}\label{eq:total-votes1}
    V_1^e(k) = \sum_{i\in\Theta} N_{i,k}\Phi_i(x_i^e(k)).
\end{equation}
The opposing candidate will then get votes of:
\begin{equation}\label{eq:total-votes2}
    V_2^e(k) = \sum_{i\in\Theta} N_{i,k}[1-\Phi_i(x_i^e(k))] = N_k - V_1^e(k),
\end{equation}
with $v_j^e(k)=V_j^e(k)/N_k$. These vote shares depend on district composition, $N_{i,k}$, election-specific ideological distances, $\mu_i^e\Rightarrow x_i^e(k)$, and voter groups' consumption preferences, $\kappa_i, \epsilon\Rightarrow x_i^e(k)$.

Our formulation, by adding an explicit group bias to \citet{dixitlondregan1996}, allows us to distinguish between descriptive and substantive representation, and also endogenizes groups' electoral influence. Redistricting affects electoral outcomes by altering district composition, $N_{i,k}$, hence changing both the relative importance of descriptive representation via votes and candidates' incentives to compete for groups' votes through transfers.

\subsection{Order of Play}\label{subsection_order-of-play}

The game proceeds as follows:
\begin{enumerate}
    \item \textbf{Redistricting:} A districting plan $\mathbf{D}=(\mathbf{N_1},...,\mathbf{N_K})$ is enacted, partitioning voters into districts.
    \item \textbf{Primary Platform Selection and Voting:} In each district, Democratic candidates announce primary-specific distributive platforms $T_{i,j,k}^1$, which imply per-capita transfers $b_{i,j,k}^1$, and voters cast ballots in the primary (closed or open).
    \item \textbf{General-Election Platform Selection and Voting:} In each district, the primary winner faces a Republican candidate and voters cast ballots based on election-specific candidate platforms $T_{i,j,k}^2$, which imply per-capita transfers $b_{i,j,k}^2$, in the general election. Given district composition, $N_{i,k}$, affinity distributions $\Phi_i$, and cutpoints $x_i^e(k)$, expected vote shares follow from $v_j^e(k)=V_j^e(k)/N_k$, and the probability that a candidate $j\in\Theta$ wins a district $k$ is
    \begin{eqnarray}
    \Psi_{mD}(k) &=& \Psi_{mD,k}^1(v^1_{mD}(k))\cdot\Psi_{mD,k}^2(v_{mD}^2(k)); \\
    \Psi_{nD}(k) &=& \Psi_{nD,k}^1(v^1_{nD}(k))\cdot\Psi_{nD,k}^2(v_{nD}^2(k)); \\
    \Psi_{R}(k) &=& 1 - \Psi_{mD}(k) - \Psi_{nD}(k) .
    \end{eqnarray}
    \item \textbf{Policy Implementation:} The elected legislature implements the budget and distributes per-capita group transfers according to the winner's platform.
\end{enumerate}
We solve for the subgame perfect Nash equilibrium by backward induction.

\subsection{Minority Welfare and Evaluating Districting Plans}\label{subsection_evaluation-districting}

We evaluate districting plans by their impact on the aggregate welfare of the minority group ($mD$). An optimal districting plan $\mathbf{D}^\ast$ solves
\begin{equation}\label{eq:objective_districting}
\mathbf{D}^\ast \in \argmax_{\mathbf{D}\in\mathbf{D}^K}
\sum_{k=1}^{K} N_{mD,k}\,
E\left[U_{mD}(k)|D\right],
\end{equation}
so that minority voters' welfare is aggregated across districts according to their population share. Expected minority voters' welfare in district $k$ is 
\begin{equation}\label{eq:model_expected-benefits}
E\left[U_{mD}(k)|D\right] = \sum_{j\in \Theta}\Psi_j(k)\left[\mu_{mD,j} + \kappa_{mD}\frac{b_{mD,j,k}^{1-\epsilon}}{1-\epsilon}\right].
\end{equation}
Here, $\Psi_j(k)$ denotes the equilibrium probability that candidate type $j$ wins the general election in district $k$, $\mu_{mD,j}$ captures minority voters' ideological affinity for candidate $j$, and $b_{mD,j,k}$ denotes the distributive benefits implemented by the winning candidate in equilibrium.\footnote{At each electoral stage, ideological competition is summarized by the relative comparison $\mu_i^e=\mu_{i,2}-\mu_{i,1}$, while realized ideological payoffs derive from the primitive $\mu_{i,j}$ and the identity of the winner.} 

This formulation highlights that districting affects minority welfare through two distinct mechanisms. First, changes in district composition alter candidates' incentives to target minority voters in the general election, operating through marginal responsiveness $\phi(\cdot)$. We refer to this as the \emph{competition channel}. Second, district composition affects candidates' probabilities of winning primary and general elections, operating through aggregate support levels $\Phi(\cdot)$ and through $\Psi_j(k)$. We refer to this as the \emph{selection channel}. 

Because candidate ideology influences marginal responsiveness $\phi(\cdot)$, and hence endogenous group influence and targeting incentives, districting shapes both channels in equilibrium. In Sections~\ref{section_substantive-representation} and~\ref{section_descriptive-representation}, we isolate the competition and selection channels, respectively. Section~\ref{section_tradeoffs-representation} then analyzes their interaction and the resulting welfare tradeoffs.

\section{Electoral Competition and Distributive Benefits}\label{subsection_equilibrium-platforms}

Before analyzing the redistricting problem, we must first determine the candidates' equilibrium platforms offered to voter groups in the respective districts. In the Dixit-Londregan framework, candidates choose group-targeted promises to maximize vote share subject to a budget constraint.

\begin{lemma}\label{lemma-platforms}
Individual benefits and shares of distributive benefits offered to voters of group $i$ by candidate $j$ in district $k$
\begin{enumerate}
\item are identical across two candidates but depend on the candidates' profiles
\begin{equation}\label{eq:individual-group-equilibrium-benefits}
b^e_{i,j,k}=\frac{\pi_i^e}{\sum_{\ell}\pi^e_{\ell}N_{\ell, k}} \text{ and }
T^e_{i,j,k}=\frac{\pi_i^e N_{i,k}}{\sum_{\ell}\pi^e_{\ell}N_{\ell, k}},
\end{equation}
where
\begin{equation}\label{eq:group-power}
    \pi_i^e = [\kappa_i\phi_i(\mu_i^e)]^{1/\epsilon} \text{ with } \phi_i(.)=\Phi'_i(.);
    \end{equation}
\item differ across electoral stages if primaries are closed and candidates can adjust platforms across stages but converge to the general-election platform; 
\item coincide across electoral stages if primaries are open.
\end{enumerate}
\end{lemma}

All proofs are in Appendix~\ref{section_appendix_proofs}. The first part is qualitatively similar to \citet{dixitlondregan1996}: In equilibrium, competing candidates adopt identical distributive platforms in the same election: $b^e_{i,1,k}=b^e_{i,2,k}$ and therefore $T^e_{i,1,k}=T^e_{i,2,k}$ for each group $i$ in district $k$. However, distributive benefits and group shares depend on i) the distribution of voters $N_{i,k}$, which we will derive when we endogenize the districting scheme $\mathbf{D}$, but unlike \cite{dixitlondregan1996}, also on ii) the group's influence to swing a specific election between two specific candidates, represented by $\pi_i^e$ and discussed below in detail. This implies that although candidates in a given election adopt the same platform, the benefits offered depend on which candidates face each other, whereas group pivotality is fixed given a particular matchup.

The other two parts address our two-stage electoral process, comprising a primary and a general election, in which the primary may be open or closed, and candidates adopt platforms tailored to different electorates. A key implication is that equilibrium distributive promises are determined in the general election, in which all voter groups participate and choose between a (minority or nonminority) Democrat and a Republican. Under closed primaries, Democratic candidates compete only for a subset of voters, but once the nominee competes against a Republican for general-election votes, equilibrium promises converge to the general-election platform; empirical evidence of such between-stage adjustment is consistent with \citet{ditellaetal2023}. Under open primaries, the relevant electorate already includes all voters, so equilibrium distributive promises coincide across the nomination and general-election stages. Accordingly, while primaries affect candidate selection and ideological benefits, distributive incentives are determined in the general election, conditional on the resulting candidate matchup.

Notably, the primary system affects both the competition and the selection channel. Primaries affect the likelihood that any candidate is elected and also affect distributive benefits, as these depend on district composition, $N_{i,k}$, and voters' heterogeneous distances to different candidates, $\mu_i^e$. Hence, a group may be more or less powerful depending on which candidates are more likely to reach the general election.

\paragraph{Components of group power.}
More detailed, group power $\pi_i^e$ is increasing in:
\begin{enumerate}
\item \textbf{Marginal utility of transfers ($\kappa_i$).} Groups with higher marginal utility from distributive benefits are more responsive to promises at the margin and receive larger equilibrium shares.
\item \textbf{Marginal electoral responsiveness at the cutoff ($\phi_i(\mu_i^e)$).} A larger density at the indifference cutoff implies that a larger mass of voters lies near the margin and is responsive to small changes in policy offers, so equilibrium platforms weight groups in proportion to this marginal decisiveness. In contrast to \cite{dixitlondregan1996}, this responsiveness is election-specific, and group pivotality depends on the specific candidate matchups. 
\item \textbf{Returns to consumption ($1/\epsilon$).} As diminishing returns intensify (higher $\epsilon$), group power differences compress, and distributive allocations become closer to proportional-to-size averages. In the limit $\epsilon\to\infty$, $\pi_i\to 1$, so $b_{i,j,k}\to B_k/N_k$ and $T_{i,j,k}\to N_{i,k}/N_k$.
\end{enumerate}

Because total transfers are fixed within a district, increases in one group's power reduce other groups' shares through the contest structure in \eqref{eq:individual-group-equilibrium-benefits}. These across-district externalities are compositional: reallocating voters to one district necessarily changes the composition and payoffs in others, providing the channel through which districting affects distributive outcomes by altering $N_{i,k}$ and, through the location of voters relative to indifference, $\phi_i^e(\mu_i^e)$ and $\pi_i^e$.

Given these equilibrium electoral outcomes, we now turn to the redistricting problem. We divide the analysis into three parts: Section~\ref{section_substantive-representation} examines per-voter distributive benefits $b_{i,j,k}$ (the ``slope'' of the utility function) to describe substantive representation by holding the electoral matchup fixed and focusing on the competition channel; Section~\ref{section_descriptive-representation} analyzes the ideological utility associated with electing different types of representatives (the ``mean'' or level of utility) to describe the descriptive representation, determining electoral matchups and the selection channel; finally, Section~\ref{section_tradeoffs-representation} combines these two to characterize the districting schemes that maximize minority voters' overall returns and highlight trade-offs between descriptive and substantive representation when group pivotality is endogenous.

\section{Redistricting and Substantive Representation}\label{section_substantive-representation}

Substantive representation in our context can be examined through the allocation of distributive benefits and their interaction with redistricting. The redistricting problem in this environment addresses two questions: how voters are distributed across districts, and how districts compare in their electoral composition. To isolate the competition channel from the selection channel, we hold the general-election matchup constant in all districts (either $\{mD, R\}$ or $\{nD, R\}$) and treat the resulting ideological distances $\mu_i^G$ as fixed parameters. Conditional on this general-election matchup, $\mu_i^G$ and hence $\pi_i^e=\pi_i(\mu_i^G)\equiv\pi_i$ group pivotality $\pi_i^e$ are constant and treated as exogenous. Hence, this specification abstracts from feedback effects operating through candidate selection and changes in matchups. In Section~\ref{section_tradeoffs-representation}, we relax this restriction and allow electoral matchups and group pivotality to adjust endogenously with district composition.

Characterizing districting schemes that provide the most benefits to minorities depends on the behavior of (\ref{eq:individual-group-equilibrium-benefits}) on the two-dimensional simplex $S^2$. We are particularly interested in its behavior on the surface $N_{mD,k}+N_{nD,k}+N_{R,k}=N_k=\mathbf{N}/K$. We rewrite (\ref{eq:individual-group-equilibrium-benefits}), minorities' distributive benefits as a group's share in a given district by any candidate,\footnote{We can ignore the candidate subscript as candidates in the same district promise the same benefits.} as
\begin{eqnarray}
T_{mD,k}=f(N_{mD,k},N_{nD,k}) &=& \frac{\pi_{mD} N_{mD,k}}{\pi_{mD} N_{mD,k} + \pi_{nD} N_{nD,k} + \pi_R N_{R,k}} \label{eq:minority-benefits-population-power-orig} \\
                &=& \frac{\pi_{mD}N_{mD,k}}{(\pi_{mD}-\pi_R)N_{mD,k}+(\pi_{nD}-\pi_R)N_{nD,k} +\pi_RN_k}\geq0,\label{eq:minority-benefits-population-power}
\end{eqnarray}
where $\Pi_k\equiv(\pi_{mD}-\pi_R)N_{mD,k}+(\pi_{nD}-\pi_R)N_{nD,k} +\pi_R N_k$ is the aggregate group power of district $k$. The derivatives with respect to a group's power are:
\begin{eqnarray}
\frac{\partial f(.)}{\partial \pi_{mD}} &=& \frac{N_{mD,k}\left(\pi_R(N_k-N_{mD,k}-N_{nD,k})+\pi_{nD}N_{nD,k} \right)}{\Pi_k^2} \geq 0;\\
\frac{\partial f(.)}{\partial \pi_{nD}} &=& -\frac{\pi_{mD}N_{mD,k}N_{nD,k}}{\Pi_k^2}\leq0;\\
\frac{\partial f(.)}{\partial \pi_{R}} &=& -\frac{\pi_{mD}N_{mD,k}(N_k-N_{mD,k}-N_{nD,k})}{\Pi_k^2}\leq0.
\end{eqnarray}
Increases in the minority group’s power are beneficial, while increases in the power of either of the other groups decrease the minority's distributive benefits. We now consider the optimal allocation of voters within and across districts.

\subsection{Districting Scheme and District Composition}

Since our baseline model assumes homogeneous preferences within each group and district, the distributive objective can be written at the group level. We seek a valid districting scheme $\mathbf{D^\ast}$ such that
\begin{equation}
    \mathbf{D^\ast} \in \argmax_{\mathbf{D}\in \mathcal{D}^K} \sum_{k=1}^{K}N_{mD,k}E\left[U_{mD}(b_{mD,k})|\mathbf{D}) \right].
\end{equation}

We evaluate the derivatives of minority voters' distributive share with respect to district populations:\footnote{The solution to the utility-maximizing districting scheme may not be unique.  Hence, we let $\mathcal{D^\ast}$ be a representative element of the set of all possible schemes.} 
\begin{eqnarray}
\frac{\partial f(.)}{\partial N_{mD,k}} &=& \frac{\pi_{mD}\left(\pi_{nD}N_{mD,k}+\pi_R(N_k-N_{mD,k})\right)}{\Pi_k^2} >0, \label{eq:foc-minority-benefits-mD-voters}\\
\frac{\partial f(.)}{\partial N_{nD,k}} &=& \frac{\pi_{mD}N_{mD,k}(\pi_R-\pi_{nD})}{\Pi_k^2} \gtreqless 0.\label{eq:foc-minority-benefits-nD-voters}
\end{eqnarray}
Adding minority voters always increases their share of benefits. However, the sign of (\ref{eq:foc-minority-benefits-nD-voters}) is ambiguous. If Republicans are politically more powerful than nonminority Democrats ($\pi_R>\pi_{nD}$), minority distributive shares rise when $nD$ voters replace $R$ voters in a district (reducing $\Pi_k$). These local comparative statics also hint that 
any districting plan with two interior districts offers opportunities to raise minority benefits: reallocating voters generates gains in one district that outweigh corresponding losses in another. Hence, we can state 
\begin{proposition}[District Composition and Pairing]\label{proposition-distributive-district-composition} 
Assume $\pi_{nD}\neq\pi_R$ and $N_{mD,k}\neq N_{mD,l}$ for $k\neq l$. Then, in any districting plan that maximizes minority distributive benefits
\begin{enumerate}
\item at most one district will lie in the interior of $S^2$;
\item districts with higher minority concentration pair minority voters with the less powerful nonminority group, while more powerful nonminority voters are concentrated elsewhere.
\end{enumerate}
\end{proposition}

\begin{figure}[t]
\centering
\footnotesize
\includegraphics[scale=.5]{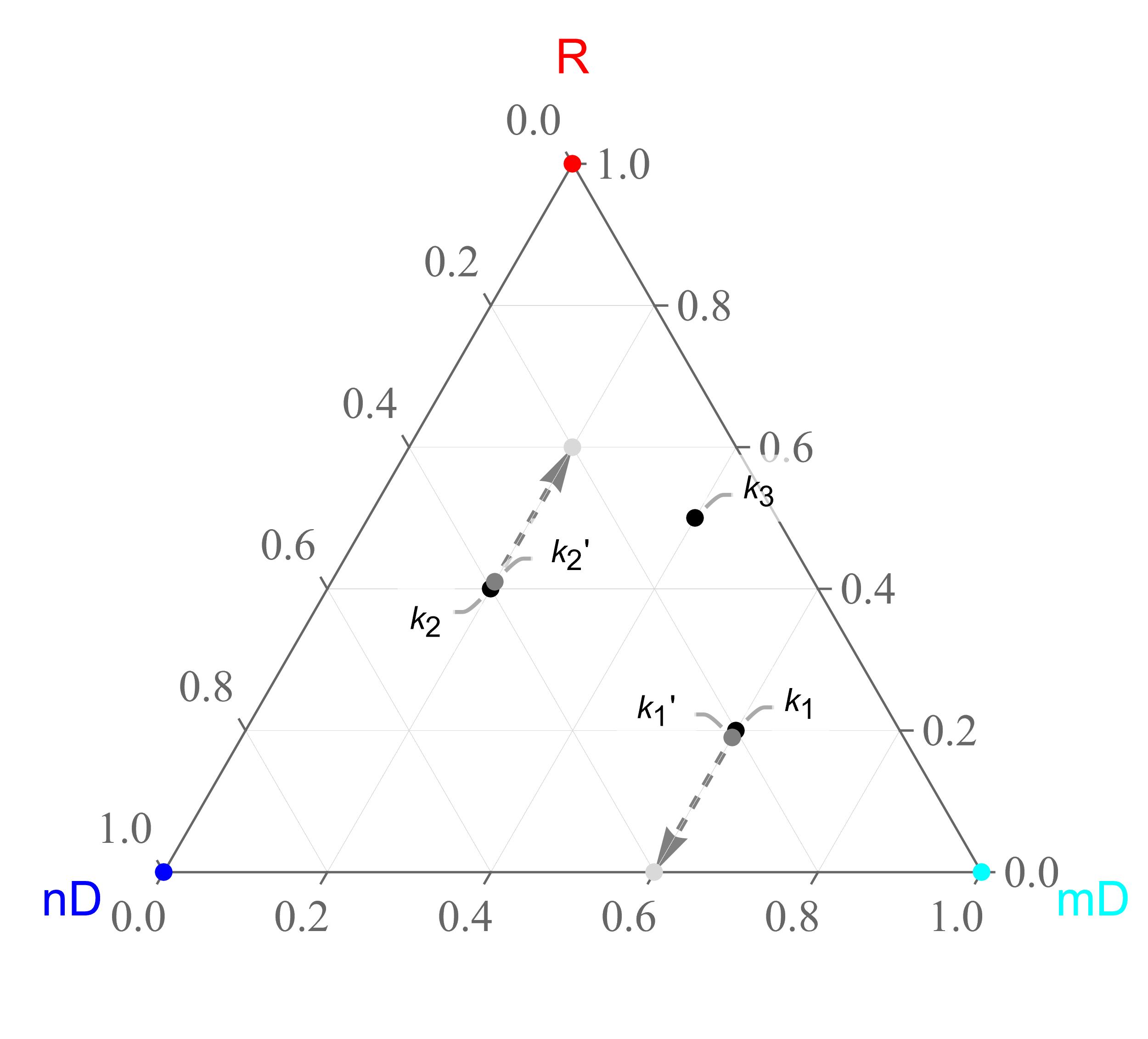}
\caption{Optimal Districting Process for $\pi_R>\pi_{nD}$ and $\Pi_1\leq\Pi_2$ with $K=3$.} 
\label{fig:Prop1}
\end{figure}

The key implication is geometric: generically, optimal districting schemes contain at most one interior district. This structure, in turn, implies a systematic pairing of minority voters with less powerful nonminority voter groups. We can illustrate the intuition of the proof with Figure~\ref{fig:Prop1}. Consider two districts $k_1$ and $k_2$ with $N_{mD,1}>N_{mD,2}$ and $\pi_R>\pi_{nD}$. Focusing on the simplex's interior, the goal is to shift voters of the more powerful nonminority group out of the minority-populated district and into the other district. As shown in Figure~\ref{fig:Prop1}, accomplishing this requires moving Republicans from $k_1$ to $k_2$ and nonminority Democrats from $k_2$ to $k_1$, where they exert greater influence. If $k_1$ is less powerful than $k_2$ ($\Pi_1\leq\Pi_2$), then minority benefits rise in $k_1$ (with more minority voters) by more than they fall in $k_2$ (with fewer). Hence, average payoffs for minority voters across districts increase, but the district's power across groups decreases. This process continues until one district reaches the simplex's border. It can then be repeated with any other pair of interior districts satisfying $N_{mD,k}\neq N_{mDl}$ and $\frac{N_{mD,k}^2}{\Pi_k^2} > \frac{N_{mDl}^2}{\Pi_l^2}$ (see districts $k_2$ and $k_3$ in Figure~\ref{fig:Prop1}). In equilibrium, at least one district is on the simplex border. 

Because $f(\cdot)$ is increasing in $N_{mD,k}$ and decreasing in the presence of a more powerful opposing group, minority voters gain when their higher-minority districts are ``paired'' with the less powerful nonminority group, packing powerful nonminority voters into other (more competitive) districts due to the groups' competition for a share of each district's fixed legislative pie. The arising across-district externalities are compositional: reallocating voters to one district necessarily changes the composition and, hence, the payoffs in others. Overall, this pattern reflects a packing-and-cracking dynamic defined by ethnicity and electoral influence rather than by partisanship.

\subsection{Voter Distribution and Minority Distributive Benefits}

\begin{figure}[t]
    \centering
\subfigure[Concave: $\pi_{mD}=10$, $\pi_{nD}=5$, $\pi_R=2$]{\label{fig:Minority-Distributive-Payoffs_concave}\includegraphics[scale=.55]{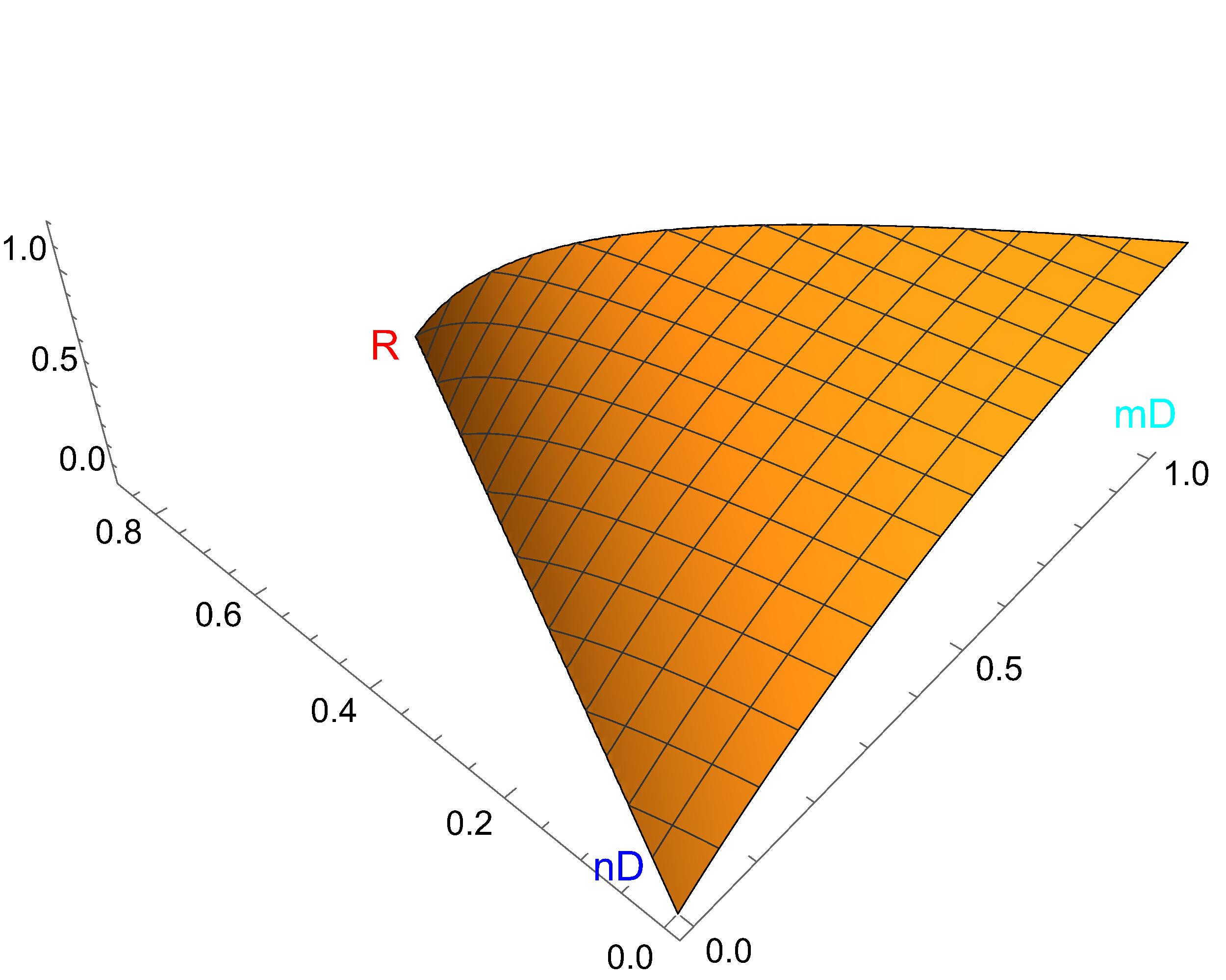}}
    \subfigure[Convex: $\pi_{mD}=2$, $\pi_{nD}=5$, $\pi_R=10$]{\label{fig:Minority-Distributive-Payoffs_convex}\includegraphics[scale=.55]{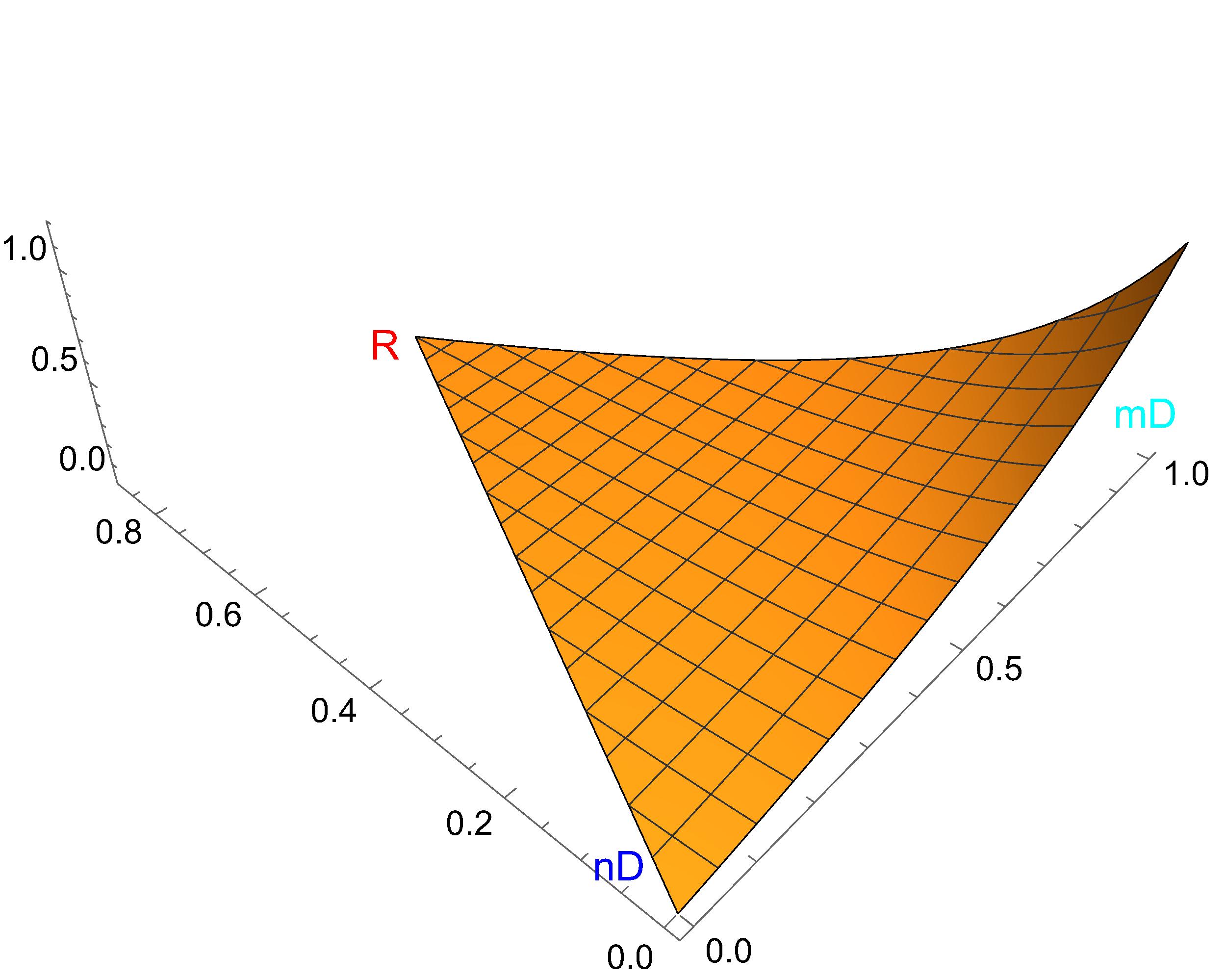}}
    \caption{Concave and Convex Minority Distributive Benefits.}
    \label{fig:Minority-Distributive-Payoffs}
\end{figure}

We next characterize how minority distributive benefits vary with district composition. To determine the voter concentration effects on distributive benefits, we evaluate the curvature of $T_{mD,k}$ with respect to district population shares, illustrated on the simplex $S^2$.\footnote{The surfeit of boundary conditions makes the usual maximization solution via Lagrange multipliers opaque. Still, we can gain insight into the solution by examining the concavity/convexity of the payoff function with respect to the number of minority voters in the district.} We calculate the determinants of the principal minors of the Hessian matrix:
\begin{equation}
H = \begin{bmatrix}
    \frac{\partial^2 f(.)}{\partial N_{mD,k}^2} & \frac{\partial^2 f(.)}{\partial N_{mD,k}N_{nD,k}} \\
    \frac{\partial^2 f(.)}{\partial N_{nD,k}N_{mD,k}} & \frac{\partial^2 f(.)}{\partial N_{nD,k}^2}
  \end{bmatrix}.
\end{equation}
The determinant of $H$ is positive on the interior for $\pi_{nD}\neq\pi_{R}$, but the value of $\frac{\partial^2 f(.)}{\partial N_{mD,k}^2}$ depends on parameter values. Figure~\ref{fig:Minority-Distributive-Payoffs} illustrates that the surface is concave when the minority group's power (electoral leverage) is larger than that of others, and convex when minority power is lower.  This forms the basis for 
\begin{lemma}[Curvature of Distributive Benefits]\label{lemma-distributive-districting-payoff-simplex}
If $\pi_{mD}=\max_{i\in\Theta}\{\pi_i\}$, then $T_{mD,k}$ is concave on the interior of $S^2$; if $\pi_{mD}=\min_{i\in\Theta}\{\pi_i\}$, then $T_{mD,k}$ is convex on the interior of $S^2$.
\end{lemma}

The curvature of the surface governs how marginal gains from concentrating minority voters change as their district share increases. When minority voters are the least powerful group (lowest $\pi_i$), marginal gains increase with concentration; when they are the most powerful group (highest $\pi_i$), marginal gains diminish with concentration. These curvature properties imply that concentration has increasing marginal returns when minority electoral power is low, and decreasing marginal returns when it is high.

\subsection{Minority Welfare Under Competition Channel}

Our results map into minority welfare. Holding the general-election matchup fixed, ideological benefits do not vary across districting plans, and differences in minority welfare arise purely through distributive benefits. Using the welfare objective from \eqref{eq:model_expected-benefits}, we can define minority welfare under the competition channel as
\begin{equation}\label{eq:minority-distributive-welfare}
W^{c}(D)= \sum_{k=1}^KN_{mD,k}\cdot\kappa_{mD}\cdot\frac{b_{mD,k}^{1-\epsilon}}{1-\epsilon},
\end{equation}
where $b_{mD,k}$ depends on district composition as described by \eqref{eq:individual-group-equilibrium-benefits}'s contest structure. We can also describe minority voter concentration by the maximum difference between the minority population of any two districts in an optimal districting scheme:
\begin{equation}
R(\mathbf{D}) = \max_{d_k, d_l \in \mathbf{D^\ast}} \left(N_{mD,k} - N_{mD,l}\right).
\end{equation}

Because of the equilibrium promises that are identical across candidates in a fixed general-election matchup, our welfare comparisons across districting plans coincide with the ranking induced by minority distributive shares as in Lemma~\ref{lemma-distributive-districting-payoff-simplex}. Since optimal values on a concave surface are less dispersed than on a convex surface, optimal gerrymanders for distributive benefits divide minority voters more equally as they gain power relative to other groups.\footnote{In the extreme convex case, optimal allocations push minority voters toward boundary districts. Conversely, when $T_{mD,k}$ is concave, and population size permits, optimal allocations yield $N_{mD,k} > 0$ for all districts.} Applying the curvature properties of Lemma~\ref{lemma-distributive-districting-payoff-simplex}, we can state

\begin{proposition}[Minority Voter Allocation and Welfare]\label{proposition_distributive-districting-distribution}
Conditional on the competition channel and holding the general-election matchup fixed, optimal districting schemes that maximize minority distributive welfare will
\begin{enumerate} 
\item concentrate minority voters in as few districts as feasible when their electoral influence is the lowest, $\pi_{mD}=\min_{i\in\Theta}\{\pi_i\}$;
\item disperse minority voters more evenly across districts when their electoral influence is the largest, $\pi_{mD}=\max_{i\in\Theta}\{\pi_i\}$.
\end{enumerate}
\end{proposition}

When minority voters have low electoral influence, because their marginal utility of distributive benefits is low ($\kappa_{mD}$ small), they are ideologically distant from the electoral indifference point ($\phi_{mD}(\mu_{mD}^G)$ small), or diminishing returns to transfers are weak, candidates face weak incentives to target them, making it optimal to concentrate minority voters into a few districts where their numbers secure larger within-district benefits. By contrast, when minority voters have high electoral influence, because they place high value on distributive benefits ($\kappa_{mD}$ high), are electorally pivotal ($\phi_{mD}(\mu_{mD}^G)$ high), and diminishing returns to transfers compress power differences ($\varepsilon$ high), minority power $\pi_{mD}$ is high, candidates have strong incentives to target them, and spreading minority voters across districts maximizes total distributive benefits by intensifying competition for their support.

Moreover, $\frac{\partial R(\mathbf{D})}{\partial \pi_{mD}} \leq 0$, so that minority voters are (weakly) spread out more evenly across districts as their power increases. Table~\ref{tbl:optimal-district-minority-benefits_minimum} illustrates optimal districting schemes for a state with three districts and state population shares of $N_{mD}=25\%$, $N_{nD}=40\%$, and $N_{R}=35\%$.\footnote{The same patterns arise when we vary the magnitude of powers, state demographics, and number of districts.} As $\pi_{mD}$ increases, the range $R(\mathbf{D})$ declines and minority distributive benefits rise. When minorities are the least powerful group, they are highly concentrated (rows highlighted in blue). By contrast, when minorities are the most powerful group, they are spread evenly across all districts (rows highlighted in orange).

\begin{table}[t]
\centering
\footnotesize

\begin{tabular}{lll|lllllllll|lll}
\toprule
\multicolumn{3}{c}{\textbf{Group Power}}     & \multicolumn{3}{c}{\textit{District 1}}       & \multicolumn{3}{c}{\textit{District 2}}       & \multicolumn{3}{c}{\textit{District 3}}   & \multicolumn{3}{l}{\textbf{Distr. Benefits}}                        \\
$\pi_{mD}$ & $\pi_{nD}$ & $\pi_{R}$ & $n_{mD1}$     & $n_{nD1}$ & $n_{R1}$ & $n_{mD2}$     & $n_{nD2}$ & $n_{R2}$ & $n_{mD3}$ & $n_{nD3}$ & $n_{R3}$ & Total & Aver. & $R(\mathbf{D})$   \\ \midrule
\rowcolor{LightCyan}1        & 3        & 1       & 75\%     & 0\%      & 25\%    & 0\%      & 64\%     & 36\%    & 0\%      & 56\%     & 44\%    & 0.750         & 0.250 & 75\% \\
2        & 3        & 1       & 44\%     & 0\%      & 56\%    & 0\%      & 100\%    & 0\%     & 31\%     & 20\%     & 49\%    & 0.974         & 0.325 & 44\% \\
3        & 3        & 1       & 39\%     & 0\%      & 61\%    & 36\%     & 20\%     & 44\%    & 0\%      & 100\%    & 0\%     & 1.167         & 0.389 & 39\% \\
4        & 3        & 1       & 0\%      & 100\%    & 0\%     & 37\%     & 20\%     & 43\%    & 38\%     & 0\%      & 62\%    & 1.300         & 0.433 & 38\% \\
5        & 3        & 1       & 30\%     & 35\%     & 35\%    & 30\%     & 0\%      & 70\%    & 15\%     & 85\%     & 0\%     & 1.426         & 0.475 & 15\% \\ \hline
\rowcolor{LightCyan}1        & 3        & 3       & 0\%      & 54\%     & 46\%    & 0\%      & 54\%     & 46\%    & 75\%     & 13\%     & 12\%    & 0.500         & 0.167 & 75\% \\
\rowcolor{LightCyan}2        & 3        & 3       & 0\%      & 53\%     & 47\%    & 0\%      & 51\%     & 49\%    & 75\%     & 15\%     & 10\%    & 0.667         & 0.222 & 75\% \\
\rowcolor{LightOrange}3        & 3        & 3       & 25\%     & 40\%     & 35\%    & 25\%     & 40\%     & 35\%    & 25\%     & 40\%     & 35\%    & 0.750         & 0.250 & 0\%  \\
\rowcolor{LightOrange}4        & 3        & 3       & 25\%     & 49\%     & 26\%    & 25\%     & 33\%     & 42\%    & 25\%     & 38\%     & 37\%    & 0.923         & 0.308 & 0\%  \\
\rowcolor{LightOrange}5        & 3        & 3       & 25\%     & 39\%     & 36\%    & 25\%     & 40\%     & 35\%    & 25\%     & 41\%     & 34\%    & 1.071         & 0.357 & 0\%  \\ \hline
\rowcolor{LightCyan}1        & 3        & 5       & 0\%      & 93\%     & 7\%     & 0\%      & 2\%      & 98\%    & 75\%     & 25\%     & 0\%     & 0.500         & 0.167 & 75\% \\
\rowcolor{LightCyan}2        & 3        & 5       & 75\%     & 25\%     & 0\%     & 0\%      & 3\%      & 97\%    & 0\%      & 92\%     & 8\%     & 0.667         & 0.222 & 75\% \\
\rowcolor{LightCyan}3        & 3        & 5       & 0\%      & 93\%     & 7\%     & 0\%      & 2\%      & 98\%    & 75\%     & 25\%     & 0\%     & 0.750         & 0.250 & 75\% \\
4        & 3        & 5       & 40\%     & 60\%     & 0\%     & 35\%     & 60\%     & 5\%     & 0\%      & 0\%      & 100\%   & 0.876         & 0.292 & 40\% \\
5        & 3        & 5       & 37\%     & 58\%     & 5\%     & 38\%     & 62\%     & 0\%     & 0\%      & 0\%      & 100\%   & 0.987         & 0.329 & 38\%            \\ \bottomrule
\end{tabular}

\caption{Districting Plans Maximizing Minority Distributive Benefits.}
\label{tbl:optimal-district-minority-benefits_minimum}
\end{table}

The next section shows that the selection channel can generate opposing curvature properties, creating potential tension between substantive and descriptive representation.

\section{Redistricting and Descriptive Representation}\label{section_descriptive-representation}

Descriptive representation can be examined by analyzing electoral outcomes and their effects on voters' ideological benefits, abstracting from the competition channel and distributive benefits. Our analysis proceeds in two steps: first, we examine the selection of minority and nonminority candidates; second, we describe the expected ideological benefits induced by electoral outcomes and characterize the optimal districting scheme that maximizes minority voters' ideological benefits.

To isolate the selection channel in this section, we assume the following ordering of voter group support, holding the general-election matchup fixed, and hence treating $\mu_i^2$, $\Phi^2_{i}(\mu_{i}^2)$, and $\pi^2_{i}(\mu_{i}^2)$ as constant. For the relevant election stage $e$ in which a minority candidate is on the ballot, minority voters are most likely to support a minority candidate, nonminority Democrats are intermediate, and Republicans are least likely to do so:\footnote{The partisan-ethnic ordering in candidate support is documented in \citet{washington2006}, \citet{mcconnaughyetal2010}, and \citet{petrowetal2018}, showing that minority voters are more likely than same-party nonminority voters to support minority candidates.}
\begin{equation}\label{eq:min-candidate-order}
\Phi^e_{mD}(\mu_{mD}^e)>\Phi^e_{nD}(\mu_{nD}^e)>\Phi^e_{R}(\mu_{R}^e).
\end{equation}
We maintain this ordering for both the primary ($e=1$) and general election ($e=2$) whenever a minority candidate is on the ballot. Similarly, nonminority Democratic voters are more likely to support nonminority Democratic candidates, whereas Republican voters are more likely to vote for Republican candidates. Lastly, we also assume that minority voters are more likely to support Democratic candidates than Republican voters in general elections, $\Phi_{mD}^e(\mu_{mD}^e)>\Phi_{R}^e(\mu_{R}^e)$ for $e=2,3$.

\subsection{Likelihood of Minority Candidate Success}\label{subsection_min-candidate-success}

We begin by analyzing how district composition affects the likelihood that a minority candidate wins the Democratic primary and then the general election. The probabilities of candidate success follow from the winning likelihood $\Psi(v)$, which maps expected vote shares into winning probabilities, and from the expected votes for each candidate, as described in \eqref{eq:total-votes1} and \eqref{eq:total-votes2}.

Evaluating the first and second derivatives of the probability that a minority candidate wins office through the primary-general election process, denoted by $\Psi_{mD}$ under closed primaries and $\tilde{\Psi}_{mD}$ under open primaries, we obtain
\begin{proposition}[District Composition and Minority Candidate Selection]\label{proposition_min-candidate-selection}
Isolating the selection channel, the following holds for minority descriptive representation with respect to minority candidate success.  
\begin{enumerate}
\item Increasing the number of minority voters in a district weakly increases the likelihood of selecting a minority candidate:
\begin{equation}
\frac{\partial\Psi_{mD,k}}{\partial N_{mD,k}} \geq 0 \text{ and } \frac{\partial\tilde{\Psi}_{mD,k}}{\partial N_{mD,k}} \geq 0;
\end{equation}
\item Increasing the number of nonminority Democratic voters in a district has ambiguous effects on selecting a minority candidate:
\begin{equation}
\frac{\partial\Psi_{mD,k}}{\partial N_{nD,k}} \gtreqless 0 \text{ and } \frac{\partial\tilde{\Psi}_{mD,k}}{\partial N_{nD,k}} \gtreqless 0
\end{equation}
due to a weakly negative primary election effect, $\frac{\partial\Psi^1_{mD,k}}{\partial N_{nD,k}}\leq 0$ and $\frac{\partial\tilde{\Psi}^1_{mD,k}}{\partial N_{nD,k}}\leq 0$, but an ambiguous general election effect, $\frac{\partial\Psi^2_{mD,k}}{\partial N_{nD,k}}\gtreqless 0$;
\item Under open primaries, replacing $R$ voters with $nD$ voters weakly increases the likelihood of selecting a minority candidate:
\begin{equation}
\left.\frac{\partial\tilde{\Psi}_{mD,k}}{\partial N_{nD,k}}\right|_{N_k,N_{mD,k}} \geq 0 \qquad \text{with } N_{R,k}=N_k-N_{mD,k}-N_{nD,k};
\end{equation}
\item Let $s_k \equiv N_{mD,k}/N_k$ with $N_k$ fixed and $t_k\equiv N_{nD,k}/(N_{nD,k}+N_{R,k})$. Minority candidate winning probabilities are convex in minority population share $s_k$ on the interior of $S^2$ under
    \begin{enumerate}
    \item Open primaries:
    \begin{equation}
    \frac{\partial^2\tilde{\Psi}_{mD,k}}{\partial s_{k}^2} > 0 ; 
    \end{equation}
    \item Closed primaries:
    \begin{equation}
    \frac{\partial^2\Psi_{mD,k}}{\partial s_{k}^2} > 0 \Longleftrightarrow \Phi_{mD}^2(\mu_{mD}^2) > \Phi_{nD}^2(\mu_{nD}^2)+\frac{1-t_k}{t_k}\Phi_{R}^2(\mu_{R}^2).\label{eq:closed-convex-ideology} 
    \end{equation}
    \end{enumerate}
\end{enumerate}

\end{proposition}

Adding minority voters to a district weakly increases the probability that a minority candidate wins the district's office. However, adding nonminority Democrats has nuanced effects: they may support Democrats in the general election but may oppose minority candidates in the primary. District composition, including substitution effects from swapping nonminority voters, also matters. The general-election effect becomes weakly positive if 
\begin{equation}\label{eq:nd_condition_general}
(\Phi_{nD}^2(\mu_{nD}^G) - \Phi_{R}^2(\mu_R^G))\,N_{R,k} > (\Phi_{mD}^2(\mu_{mD}^G) - \Phi_{nD}^2(\mu_{nD}^G))\,N_{mD,k} \text{ } \Rightarrow \text{ } \frac{\partial\Psi^2_{mD,k}}{\partial N_{nD,k}} > 0
\end{equation}
and $N_{R,k}>0$, which would still imply ambiguity on the overall likelihood of minority success, $\frac{\partial\Psi_{mD,k}}{\partial N_{nD,k}} \gtreqless 0$  and $\frac{\partial\tilde{\Psi}_{mD,k}}{\partial N_{nD,k}} \gtreqless 0$, and illustrate opposing effects between the primary election and the general election. In contrast, if the condition is reversed, both the primary and general elections would exhibit negative effects of increasing the number of nonminority voters on minority candidates' chances. When nonminority Democrats replace Republicans rather than minority voters, the general-election effect is positive and offsets part of the negative primary effect. The overall effect is therefore ambiguous. When nonminority Democrats instead replace minority voters, both effects are negative, and minority electoral success falls unambiguously. In open primaries, the effect is weakly positive because replacing Republicans increases minority electoral success in both the primary and the general election.

The convexity of the winning probability implies that districting schemes that maximize descriptive minority representation concentrate minority voters in as few districts as possible. Under open primaries, increasing the minority share raises the probability that a minority candidate wins both stages at a linear rate, so gains compound across the two-stage process. Under closed primaries, the primary-stage probability is concave in the minority share, while the general-election probability is linear. The overall winning probability is convex when gains in the general election outweigh diminishing returns in the primary.

Although the minority voter share $s_k$ affects both primary and general-election probabilities, the sign of curvature under closed primaries depends only on the relative composition of nonminority voters, $t_k = N_{nD,k}/(N_{nD,k}+N_{R,k})$. Packing benefits minority candidates when the nonminority electorate is predominantly Democratic (high $t_k$), but dispersion is preferable when the nonminority electorate is predominantly Republican (low $t_k$). Furthermore, when minority voters provide strong aggregate support for minority candidates (high $\Phi_{mD}^2(\mu_{mD}^G)$) or minority candidates face strong nonminority voter opposition (low $\Phi_{nD}^2(\mu_{nD}^G)$ and/or $\Phi_{R}^2(\mu_{R}^G)$), they and minority candidates benefit from packing. In contrast, when minority voters provide less electoral support (low $\Phi_{mD}^2(\mu_{mD}^G)$) or minority candidates face less nonminority voter opposition (high $\Phi_{nD}^2(\mu_{nD}^G)$ and/or $\Phi_{R}^2(\mu_{R}^G)$), minority candidates and voters benefit from dispersion.

\subsection{Voter Distribution and Minority Ideological Benefits}\label{subsection_min_ideological_benefits}

Using the winning probabilities from Section~\ref{subsection_min-candidate-success}, we characterize minority voters' expected ideological benefits across districting schemes. We defined $\mu_i^j$ as voter group $i$’s mean ideological benefit for candidate type $j$. Normalizing minority voters’ ideological payoffs as $\mu_{mD}^{mD}=1$, $\mu_{mD}^{nD}=\beta$ with $\beta\in[0,1]$, and $\mu_{mD}^{R}=0$, minority voters' expected ideological benefit in district $k$ under districting scheme $D$ is
\begin{equation}
E\left[\mu_{mD}\mid k\right] = \sum_{j\in\Theta}\Psi_{j,k}\cdot\mu_{mD}^j = \begin{cases}
\Psi_{mD,k} + \beta\,\Psi_{nD,k} \qquad\text{(closed primaries)} \\
\tilde{\Psi}_{mD,k} + \beta\,\tilde{\Psi}_{nD,k} \qquad\text{(open primaries)}.
\end{cases}
\label{eq_exp-min-utility}
\end{equation}

It is natural to ask whether the districting schemes that maximize minority voters’ expected ideological benefits also maximize the likelihood of electing minority representatives. First, we evaluate the curvature of minority voters' expected ideological benefits and then its implications for packing or cracking minority voters across districts. 

\begin{lemma}[Curvature of Ideological Benefits]\label{lemma_minority-ideology-benefits}
Conditional on the selection channel, holding the general-election environments fixed, and under the normalization $\beta\in[0,1]$, there exists a unique real root $\bar{\beta}$ of the affine curvature condition.
\begin{enumerate}
\item Closed primaries. Let
\begin{equation}
\bar{\beta}^C \equiv \frac{\big(\Phi_{mD}^2(\mu_{mD}^2)-\Phi_{nD}^2(\mu_{nD}^2)\big)N_{nD,k}-\Phi_R^2(\mu_R^2)N_k}{\left(\Phi_{mD}^3(\mu_{mD}^3)-\Phi_{nD}^3(\mu_{nD}^3)\right)N_{nD,k}-\Phi_R^3(\mu_R^3)N_k}.
\end{equation}
If the denominator is positive, convexity holds iff $\beta<\bar{\beta}^C$; if negative, convexity holds iff
$\beta>\bar{\beta}^C$.
\item Open primaries. Let 
\begin{equation}
\bar{\beta}^O \equiv \frac{ \Phi_{mD}^2(\mu_{mD}^2)-\Phi_{R}^2(\mu_{R}^2)}{\Phi_{mD}^3(\mu_{mD}^3)-\Phi_{R}^3(\mu_{R}^3)}.
\end{equation}
and convexity holds iff $\beta<\bar{\beta}^O$.
\end{enumerate}
\end{lemma}

Convexity of minority voters' expected ideological benefits depends on two related considerations: how the nonminority Democratic nominee performs in the general election, and whether switching to a minority nominee improves Democratic competitiveness. Under closed primaries, these effects are captured by the structure of the $nD-R$ general election. The support differential $\Phi_{mD}^3(\mu_{mD}^3)-\Phi_{nD}^3(\mu_{nD}^3)$ measures differential Democratic cohesion across voter groups, while $\Phi_R^3(\mu_R^3)$ captures Republican crossover toward the Democratic nominee. When Democratic success in the $nD-R$ election relies primarily on intra-party cohesion rather than crossover (positive denominator of $\bar{\beta}^C$), nominating a minority candidate can strengthen general-election prospects. In this case, if the minority nominee is sufficiently competitive relative to the nonminority nominee ($\bar{\beta}^C>1$), concentration increases both descriptive representation and Democratic victory, yielding convex ideological returns. If the minority nominee is less competitive ($0<\bar{\beta}^C<1$), concentration improves ideological welfare only when minority voters place sufficient weight on descriptive representation (low $\beta$). By contrast, when Republican crossover drives Democratic success (negative denominator), nominee identity has limited partisan consequences, and concentration affects welfare primarily through descriptive representation, making convexity less likely within the relevant range of preferences.

Under open primaries, curvature depends solely on the relative general-election competitiveness of the minority and nonminority Democratic candidates. The threshold $\beta<\bar{\beta}^O$ equals the ratio of the minority candidate’s advantage over the Republican to that of the nonminority candidate. When $\beta<\bar{\beta}^O>1$, the minority nominee is more electable, and concentration increases both descriptive and partisan success. When $0<\beta<\bar{\beta}^O<1$, concentration is optimal only if voters place sufficiently high weight on descriptive representation (low $\beta^O$).

In both systems, therefore, curvature is governed by whether nominating the minority candidate materially improves general-election prospects relative to the partisan baseline.

\subsection{Minority Welfare Under Selection Channel}

We can now turn to minority welfare. Using the welfare objective from \eqref{eq:model_expected-benefits}, we can define minority welfare under the selection channel as
\begin{equation}
W^{s}(D)= \sum_{k=1}^KN_{mD,k}\cdot E[\mu_{mD}|k].
\end{equation}
Because the curvature of $E[\mu_{mD}|k]$ governs whether marginal gains from concentrating minority voters rise or fall with $N_{mD,k}$, Lemma~\ref{lemma_minority-ideology-benefits}  implies an allocation result: holding the statewide minority population fixed, convexity implies that more unequal minority distributions (packing) increase welfare, whereas concavity favors more even distributions (dispersion).

\begin{proposition}[Minority Voter Allocation and Welfare]\label{proposition_ideological-districting-distribution}
Conditional on the selection channel and holding general-election environments fixed:

\begin{enumerate}
\item Under closed primaries, concentrating minority voters weakly increases $W^s(D)$ if and only if curvature is convex. Convexity occurs when $\beta$ lies on the convex side of $\bar{\beta}^C$, with the direction determined by the sign of $\Delta_k^C$ (the denominator of $\bar{\beta}^C$ in Lemma~\ref{lemma_minority-ideology-benefits}).
\item Under open primaries, concentrating minority voters weakly increases $W^s(D)$ if and only if $\beta<\bar{\beta}^O$.
\end{enumerate}
\end{proposition}

When $\beta$ lies on the convex side of the relevant threshold, minority voters place sufficiently high weight on descriptive representation relative to partisan outcomes (low or high $\beta$, depending on the sign of $\Delta_k^C$ under closed primaries). Under closed primaries, the direction of the cutoff depends on whether Democratic success in the $nD$–$R$ election is driven primarily by differential Democratic cohesion or by Republican crossover (sign of $\Delta_k^C$). If $\Delta_k^C>0$, convexity arises when $\beta<\bar{\beta}^C$; if $\Delta_k^C<0$, convexity arises when $\beta>\bar{\beta}^C$. In this region, increasing minority concentration raises the probability of nominating a minority candidate (higher $N_{mD,k}$ in the primary stage) and, when nominee identity materially alters general-election competitiveness (large support differentials $(\Phi_{mD}^2-\Phi_{nD}^2)$ and $(\Phi_{mD}^3-\Phi_{nD}^3)$ relative to Republican crossover $\Phi_R^2,\Phi_R^3$), also increases the likelihood that the Democratic candidate defeats the Republican. Descriptive and partisan success therefore move together, yielding increasing marginal ideological returns to concentration.

When $\beta$ lies on the concave side of the threshold, candidate identity has weaker implications for aggregate partisan success. Increasing minority concentration then raises descriptive representation but has a limited effect on overall Democratic victory rates. In this case, spreading minority voters across districts can reduce aggregate Republican victories and increase total minority ideological benefits, even if fewer districts elect minority representatives.

\paragraph{Role of Primary System}

The comparison between $\bar{\beta}^C$ and $\bar{\beta}^O$ clarifies how primary institutions shape packing incentives. Under open primaries, the curvature threshold depends solely on the relative general-election competitiveness of minority and nonminority Democratic nominees. Under closed primaries, by contrast, the threshold additionally reflects how minority concentration alters primary outcomes and thereby changes the general-election environment. In particular, the sign and magnitude of $\Delta_k^C$ determine whether Democratic success in an $nD$–$R$ contest is driven primarily by intra-party cohesion or by Republican crossover. As a result, the ordering of $\bar{\beta}^C$ and $\bar{\beta}^O$ is not universal. Primary rules, therefore, affect not only who is nominated, but also whether descriptive and partisan success reinforce or offset each other in determining optimal districting. 

We now analyze the general equilibrium in which district composition jointly shapes nomination probabilities, general-election competitiveness, and group power. Primary rules determine not only who is nominated but also how minority concentration feeds back into electoral success and policy benefits.

\section{Optimal Redistricting and Tradeoffs}\label{section_tradeoffs-representation}

Sections~\ref{section_substantive-representation} and \ref{section_descriptive-representation} isolated the competition and selection channels under fixed general-election matchups. We now allow matchups and group power to adjust endogenously with district composition. This general-equilibrium feedback introduces an additional curvature component in the minority welfare function that can reinforce, attenuate, or overturn the packing incentives implied by each channel in isolation.

The key insight is that minority concentration affects welfare through a dynamic chain: changes in district composition alter nomination probabilities, which in turn reshape general-election environments and ideological matchups, thereby shifting voter responsiveness and group power, $\pi_i^e=[\kappa_i \phi_i(\mu_i^e)]^{1/\epsilon}$. Through this mechanism, districting affects both electoral selection and distributive targeting, and, critically, their interaction via aggregate support ($\Phi_i^e(\mu_i^e)$) and marginal responsiveness ($\phi_i(\mu_i^e)$). Optimal redistricting, therefore, depends on how this endogenous feedback modifies the benchmark curvature generated by the isolated channels. The benchmark determines when descriptive and substantive representation align or diverge. Endogenous feedback can preserve that alignment, resolve a divergence, or overturn it, thereby generating nonmonotonicity in minority welfare with voter concentration.

\subsection{Welfare and Channel Interaction}

We analyze minority welfare and the interaction of the two channels by rewriting our previous welfare description of \eqref{eq:objective_districting} and \eqref{eq:model_expected-benefits} to
\begin{equation}
W(D) = \underbrace{\sum_{k=1}^KN_{mD,k}\cdot \sum_{j\in \Theta}\Psi_j(k)\cdot\mu_{mD,j}}_{W^{s\ast}(D)} + \underbrace{\sum_{k=1}^KN_{mD,k}\cdot \sum_{j\in \Theta}\Psi_j(k)\cdot\kappa_{mD}\cdot\frac{b_{mD,j,k}^{1-\epsilon}}{1-\epsilon}}_{W^{c\ast}(D)}.
\end{equation}
The first term captures the selection channel through winning probabilities $\Psi_j(k)$ and ideological benefits $\mu_{mD,j}$. The second term captures the competition channel through equilibrium distributive benefits $b_{mD,j,k}$.

The two channels interact because district composition shapes winning probabilities $\Psi_j(k)$, which determine general-election matchups $\mu_i^e$ and hence voter responsiveness $\phi_i(\mu_i^e)$. Since group power satisfies $\pi_i^e=[\kappa_i\phi_i(\mu_i^e)]^{1/\epsilon}$, changes in matchups feed back into equilibrium targeting incentives:
\begin{equation}\label{eq:detailed-mechanism}
\mathbf{D} \rightarrow \{N_{i,k}\} \rightarrow s_k \rightarrow \underbrace{\Psi_j(k)}_{\text{selection}} \rightarrow  \mu_i^e \rightarrow \Phi_i^e(\mu_i^e) \text{ and } \phi_i(\mu_i^e) \rightarrow  \pi_i^e \rightarrow \underbrace{b_{i,j,k}}_{\text{competition}}.
\end{equation}

We can now evaluate how districting schemes that concentrate or disperse minority voters affect minority welfare by examining the curvature of $W(D)$ with respect to the minority share $s_k \equiv N_{mD,k}/N_k$ in district $k$, holding $N_k$ fixed. Differentiating twice, we have
\begin{equation}\label{eq:second-derivative-min-welfare-total}
\frac{\partial^2 W(D)}{\partial s_k^2} = \underbrace{\frac{\partial^2 W^s(D)}{\partial s_k^2}}_{\text{Selection - Section \ref{section_descriptive-representation}}} +\underbrace{\left.\frac{\partial^2 W^c(D)}{\partial s_k^2}\right|_{\substack{\text{fixed}\\\text{matchup}}}}_{\text{Competition - Section \ref{section_substantive-representation}}} + \underbrace{I_k}_{\text{eq. interaction}}. 
\end{equation}
The first two terms capture the direct selection and competition effects identified in Sections~\ref{section_descriptive-representation} and~\ref{section_substantive-representation}. The interaction term $I_k$ captures the additional curvature terms arising because both winning probabilities and equilibrium transfers depend on district composition through endogenous matchups and group power. At an interior allocation, positive (negative) curvature at $s_k$ implies that marginally reallocating minority voters via mean-preserving reallocations toward (away from) district $k$ locally raises minority welfare.

\paragraph{Mechanism of Interaction}
The magnitude of $I_k$ depends on how district composition affects candidate matchups and, through them, voter responsiveness and pivotality. Abstracting from constants, its structure can be summarized schematically as
\begin{equation}\label{eq:Ik-schematic}
I_k \sim \underbrace{\sum_{j\in\Theta}\frac{\partial \Psi_j(k)}{\partial s_k}\cdot \kappa_{mD}\cdot\frac{b_{mD,j,k}^{1-\epsilon}}{1-\epsilon}}_{\text{probability reweighting}}
+
\underbrace{\sum_{j\in\Theta}\Psi_j(k)\cdot\frac{\partial \kappa_{mD}\cdot\frac{b_{mD,j,k}^{1-\epsilon}}{1-\epsilon}}{\partial b}\cdot\frac{\partial b_{mD,j,k}}{\partial \pi_{mD}^e}\cdot\frac{\partial \pi_{mD}^e}{\partial s_k}}_{\text{endogenous pivotality/targeting}}
\end{equation}
with $\partial \pi_i^e/\partial s_k = \frac{\pi_i^e}{\epsilon}\frac{\phi_i'(\mu_i^e)}{\phi_i(\mu_i^e)}\frac{\partial \mu_i^e}{\partial s_k}$. The first component captures probability reweighting: increasing $s_k$ shifts weight across electoral outcomes that deliver different minority distributive benefits. The second component captures endogenous targeting: changes in $s_k$ alter matchups $\mu_i^e$, thereby shifting voter responsiveness and group power $\pi_i^e$ and modifying equilibrium transfers. Under fixed matchups, $\mu_i^e=\mu_i^G$ is locally constant, so $\partial \mu_i^e/\partial s_k=\partial\pi_i^e/\partial s_k=0$ and the endogenous pivotality/targeting component is shut down. Probability reweighting through $\partial\Psi_j(k)/\partial s_k$ is part of the selection channel and is captured by $W_{ss}^s$, while $I_k$ collects the additional curvature terms that arise once $\partial \Psi_j(k)/\partial s_k$ and $\pi_i^e$ are allowed to respond to $s_k$. Interaction effects are therefore large when (i) winning probabilities are sensitive to district composition, and (ii) matchup shifts substantially alter responsiveness elasticities and group power. When these sensitivities are small, interaction effects are second-order; when they are large, feedback can dominate the direct curvature components. We provide the formal derivation of \eqref{eq:Ik-schematic} with its primitives in Appendix~\ref{appendix_minority-welfare-curvature-details}. The sign and magnitude of $I_k$ determine whether endogenous feedback reinforces, weakens, or overturns the benchmark curvature of $W(D)$ implied by the isolated channels.

\paragraph{Isolated Channels} To summarize the isolated-channel results from Sections~\ref{section_substantive-representation} and~\ref{section_descriptive-representation}, Table~\ref{tbl:summary-channels-voter-distribution-detailed} reviews the effects of minority power and preference weights on concentrating or spreading minority voters, recognizing the effects of the different primary systems. In the following, we will compare these partial effects to the general equilibrium effects of endogenous feedback from selection to targeting incentives, illustrated by the impact of interaction effects $I_k$. That is, equilibrium feedback may reverse whether concentration or dispersion is locally welfare-improving.

\begin{table}[t]
\centering
\footnotesize
\begin{tabular}{lcccc}
\toprule
\multicolumn{5}{c}{\textbf{Closed Primaries with $\Delta_k^C>0$, Open Primaries} -- Convexity when $\beta<\bar{\beta}$} \\
\midrule
Minority Power & Preference Weight & Competition & Selection & Combined \\
\midrule
Lowest, $\pi_{mD}=\min_{i\in\Theta}\{\pi_i\}$ & $\beta<\bar{\beta}$ & Concentrate & Concentrate & \textbf{Concentration} \\
Lowest, $\pi_{mD}=\min_{i\in\Theta}\{\pi_i\}$ & $\beta<\bar{\beta}$ & Concentrate & Spread      & \textit{Divergence} \\
Highest, $\pi_{mD}=\max_{i\in\Theta}\{\pi_i\}$ & $\beta<\bar{\beta}$ & Spread      & Concentrate & \textit{Divergence} \\
Highest, $\pi_{mD}=\max_{i\in\Theta}\{\pi_i\}$ & $\beta>\bar{\beta}$ & Spread      & Spread      & \textbf{Dispersion} \\
\midrule
\multicolumn{5}{c}{} \\
\multicolumn{5}{c}{\textbf{Closed Primaries with $\Delta_k^C<0$} -- Convexity when $\beta>\bar{\beta}$} \\
\midrule
Minority Power & Preference Weight & Competition & Selection & Combined \\
\midrule
Lowest, $\pi_{mD}=\min_{i\in\Theta}\{\pi_i\}$ & $\beta<\bar{\beta}$ & Concentrate & Spread      & \textit{Divergence} \\
Lowest, $\pi_{mD}=\min_{i\in\Theta}\{\pi_i\}$ & $\beta>\bar{\beta}$ & Concentrate & Concentrate & \textbf{Concentration} \\
Highest, $\pi_{mD}=\max_{i\in\Theta}\{\pi_i\}$ & $\beta<\bar{\beta}$ & Spread      & Spread      & \textbf{Dispersion} \\
Highest, $\pi_{mD}=\max_{i\in\Theta}\{\pi_i\}$ & $\beta>\bar{\beta}$ & Spread      & Concentrate & \textit{Divergence} \\
\bottomrule
\end{tabular}
\caption{Isolated Channels with Fixed Matchups and Minority Voter Distribution.}
\label{tbl:summary-channels-voter-distribution-detailed}
\end{table}

The decomposition clarifies when descriptive and substantive representation align or diverge. When the direct curvature components share the same sign, the channels align; when their signs differ, or interaction effects dominate, trade-offs arise.

\begin{definition}[Channel Alignment, Divergence, and Overturning]
\label{definition-alignment-overturning}
Let 
\[
W_{ss}^s \equiv \frac{\partial^2 W^s(D)}{\partial s_k^2},
\qquad
W_{ss}^{c,0}\equiv \left.\frac{\partial^2 W^c(D)}{\partial s_k^2}\right|_{\substack{\text{fixed}\\\text{matchup}}},
\qquad
C_k \equiv W_{ss}^s + W_{ss}^{c,0}.
\]
\begin{enumerate}
\item The selection and competition channels are \emph{locally aligned} at $s_k$ if $\sign(W_{ss}^s)=\sign(W_{ss}^{c,0})$.
\item The channels \emph{diverge locally} at $s_k$ if $\sign(W_{ss}^s)\neq\sign(W_{ss}^{c,0})$.
\item Equilibrium feedback \emph{overturns} the fixed-matchup benchmark at $s_k$ if $\sign(C_k+I_k)\neq \sign(C_k)$.
\end{enumerate}
\end{definition}

\subsection{Alignment and Divergence of Channels}

Total minority welfare curvature satisfies
\begin{equation}
W_{ss}(D) = C_k + I_k,
\qquad
C_k \equiv W_{ss}^s + W_{ss}^{c,0}.
\end{equation}
The sign of $C_k$ determines whether the selection and competition channels align or diverge under fixed matchups, while the interaction term $I_k$ captures endogenous feedback through electoral sensitivity and targeting incentives. For the curvature of the selection channel (Table~\ref{tbl:summary-channels-voter-distribution-detailed}), we state
\begin{definition}[Selection Curvature Regions]\label{definition_selection_regions}
Fix district $k$. The selection channel curvature in $s_k$ partitions $\beta$ into a convex region and a concave region as follows.
\begin{enumerate}
\item Open primaries. The selection channel is \emph{selection-convex} iff $\beta < \bar{\beta}^{O}$ and \emph{selection-concave} iff $\beta > \bar{\beta}^{O}$.
\item Closed primaries with $\Delta_k^{C}>0$. The selection channel is \emph{selection-convex} iff $\beta < \bar{\beta}^{C}$ and \emph{selection-concave} iff $\beta > \bar{\beta}^{C}$.
\item Closed primaries with $\Delta_k^{C}<0$. The selection channel is \emph{selection-convex} iff $\beta > \bar{\beta}^{C}$ and \emph{selection-concave} iff $\beta < \bar{\beta}^{C}$.
\end{enumerate}
\end{definition}

Given the potential magnitude of the interaction term $I_k$, we need to assess how electorally stable districts are to small changes in district composition. In some districts, winning probabilities vary gradually with minority concentration, and ideological matchups shift only modestly. Consequently, endogenous group power adjusts only marginally in response to compositional changes. In contrast, in some districts, small changes in minority concentration generate large shifts in candidate selection ($\Psi_j(k)$), ideological matchups ($\mu_i^e$), or marginal voter densities ($\phi_i(\mu_i^e)$). We refer to these districts as either \emph{safe} or \emph{tipping}.\footnote{See Appendix~\ref{appendix_minority-welfare-curvature-details} for derivative-based sufficient conditions.}

\begin{definition}[Safe Districts]\label{definition-safe-districts}
District $k$ is \emph{locally safe} at $s_k$ if $\partial \Psi_j(k)/\partial s_k$, $\partial \mu_i^e/\partial s_k$, and $\phi_i'(\mu_i^e)/\phi_i(\mu_i^e)$ are all uniformly small in a neighborhood of $s_k$.
\end{definition}

\begin{definition}[Tipping Districts]\label{definition-tipping-districts}
District $k$ is \emph{locally tipping} at $s_k$ if at least one of $\partial \Psi_j(k)/\partial s_k$, $\partial \mu_i^e/\partial s_k$, or $\phi_i'(\mu_i^e)/\phi_i(\mu_i^e)$ is large in magnitude in a neighborhood of $s_k$, so that equilibrium feedback need not be second-order relative to $C_k$.
\end{definition}

Tipping does not imply discontinuities in welfare; rather, it corresponds to regions where the derivatives governing electoral outcomes and pivotality are large in magnitude. In such environments, the interaction term $I_k$ can be first-order relative to $C_k$; equilibrium feedback may reinforce or overturn the fixed-matchup benchmark. Considering the curvature effects of the isolated channels, as well as the interaction effects in different electoral environments, we predict 

\begin{lemma}[Equilibrium Alignment and Divergence]\label{lemma_general_equilibrium}
Let $C_k = W_{ss}^s + W_{ss}^{c,0}$ denote the fixed-matchup curvature benchmark at $s_k$, and let $I_k$ denote the interaction term. Then total curvature satisfies
\begin{equation}
W_{ss}(D) = C_k + I_k.
\end{equation}

\begin{enumerate}
\item Fixed-Matchup Benchmark. If $I_k=0$, alignment arises when $\sign(W_{ss}^s)=\sign(W_{ss}^{c,0})$, and divergence arises when $\sign(W_{ss}^s)\neq\sign(W_{ss}^{c,0})$.

\item Safe Districts. If district $k$ is \emph{locally safe} and $|C_k|$ is bounded away from zero, then $|I_k|$ is second-order and $\sign(C_k+I_k)=\sign(C_k)$. Hence, alignment or divergence under fixed matchups is locally preserved.

\item Tipping Districts. If district $k$ is \emph{locally tipping}, then $I_k$ may be first-order relative to $C_k$. In this case:
    \begin{enumerate}
    \item If $\sign(I_k)=\sign(C_k)$, equilibrium feedback reinforces the fixed-matchup benchmark;
    \item If $\sign(I_k)\neq\sign(C_k)$ and $|I_k|>|C_k|$, equilibrium feedback overturns the fixed-matchup benchmark.
    \end{enumerate}
\end{enumerate}
\end{lemma}

Under fixed matchups ($I_k=0$), alignment and divergence reduce to the benchmark $C_k$ and follow directly from Propositions~\ref{proposition_distributive-districting-distribution} and~\ref{proposition_ideological-districting-distribution}.

In safe districts, composition changes do not meaningfully change who runs or wins. Districting mainly shifts the size of the represented minority constituency rather than the electoral environment that determines targeting; endogenous changes in group power remain second-order relative to the direct curvature components. As a result, the alignment established in the isolated model is preserved under general equilibrium.

In contrast to safe districts, tipping districts correspond to competitive electoral environments in which small changes in district composition can flip the general election or alter primary outcomes. In such districts, large values of $\partial \Psi_j(k)/\partial s_k$, $\partial \mu_i^e/\partial s_k$, or $\phi_i'(\mu_i^e)/\phi_i(\mu_i^e)$ can generate first-order movements in group power and targeting incentives, so that endogenous feedback may reinforce or eliminate the trade-off between descriptive and substantive representation. In divergence regions, tipping districts may therefore resolve the trade-off by reversing the sign of total curvature, effectively restoring alignment under general equilibrium.

Before turning to voter allocation, we summarize our geometric equilibrium findings in Table~\ref{tbl:summary-primitives-interaction}.

\begin{table}[t]
\centering
\scriptsize
\begin{tabular}{lcccc}
\toprule
\multicolumn{5}{c}{\textbf{Open Primaries and Closed Primaries with $\Delta_k^C>0$}} \\
\midrule
Distributive Leverage & Selection Environment & Competition & Selection & Interaction  \\
\midrule
Low $\kappa_{mD}$ or $\phi_{mD}(.)$ 
& $\beta$ in convex region & Concentrate & Concentrate & Reinforce (if stable or $I_k>0$)  \\
Low $\kappa_{mD}$ or $\phi_{mD}(.)$ & $\beta$ in concave region & Concentrate & Spread & May Resolve (if tipping) \\
High $\kappa_{mD}$ and $\phi_{mD}(.)$ & $\beta$ in convex region & Spread & Concentrate & May Resolve (if tipping) \\
High $\kappa_{mD}$ and $\phi_{mD}(.)$ & $\beta$ in concave region & Spread & Spread & Reinforce (if stable or $I_k<0$) \\
\midrule
\multicolumn{5}{c}{} \\
\multicolumn{5}{c}{\textbf{Closed Primaries with $\Delta_k^C<0$}} \\
\midrule
Distributive Leverage & Selection Environment & Competition & Selection & Interaction  \\
\midrule
Low $\kappa_{mD}$ or $\phi_{mD}(.)$ & $\beta$ in concave region & Concentrate & Concentrate & Reinforce (if stable or $I_k>0$) \\
Low $\kappa_{mD}$ or $\phi_{mD}(.)$ & $\beta$ in convex region & Concentrate & Spread & May Resolve (if tipping) \\
High $\kappa_{mD}$ and $\phi_{mD}(.)$ & $\beta$ in concave region & Spread & Spread & Reinforce (if stable or $I_k<0$)  \\
High $\kappa_{mD}$ and $\phi_{mD(.)}$ & $\beta$ in convex region & Spread & Concentrate & May Resolve (if tipping)  \\
\bottomrule
\end{tabular}
\caption{Fixed-Matchup Channels and General-Equilibrium Interaction.}
\label{tbl:summary-primitives-interaction}
\end{table}

\subsection{Minority Welfare and Voter Distribution}

Lemma~\ref{lemma_general_equilibrium} characterizes the curvature of total minority welfare as
\begin{equation}
W_{ss}(D) = W^s_{ss} + W^{c,0}_{ss} + I_k= C_k + I_k,
\end{equation}
where $C_k$ captures the fixed-matchup benchmark and $I_k$ captures endogenous electoral feedback. This decomposition implies that nonmonotonic minority welfare can either arise from divergence between descriptive and substantive representation (benchmark $C_k$) or from overturning feedback in competitive districts (interaction $I_k$).

Endogenous feedback is economically relevant only when changes in minority concentration alter the general-election environment (Definition~\ref{definition-tipping-districts}, Lemma~\ref{lemma_general_equilibrium}). At low minority shares, marginal increases in $s_k$ do not affect the $nD$–$R$ matchup; at high shares, they do not alter the $mD$–$R$ matchup. Only at intermediate shares can marginal changes in $s_k$ shift nomination probabilities and general-election matchups, generating first-order changes in pivotal densities and group power. For example, moving from $s_k=0.1$ to $s_k=0.12$ in a safe district barely changes $\Psi$ or $\mu^e$, whereas the same change near a 50–50 cutoff can flip the likely nominee and the ensuing general-election matchup. This localization follows the standard single-index probabilistic voting structure \citep{lindbeckweibull1987,dixitlondregan1996,perssontabellini2001}, in which electoral success is a smooth function of a voter utility index and the marginal electoral sensitivity $\partial \Psi_j(k)/\partial s_k$ is largest in a neighborhood of the unique electoral cutoff.

\begin{lemma}[Tipping Region]\label{lemma_tipping_region}
Suppose electoral success probabilities $\Psi_j(k)$ admit a single-index representation with a unique cutoff and endogenous objects in \eqref{eq:detailed-mechanism} enter $I_k$ only through \eqref{eq:Ik-schematic}. Then there exists a connected interval $(\underline{s},\overline{s})\subset(0,1)$ such that $I_k$ is first-order only on $(\underline{s},\overline{s})$ and second-order elsewhere.
\end{lemma}

The endogenous feedback loop from \eqref{eq:detailed-mechanism} is only at play when composition changes reweight electoral outcomes or move the general-election environment. Under a standard single-index probabilistic voting structure, this occurs only near the unique cutoff. Hence, the probability-reweighting and pivotality terms in \eqref{eq:Ik-schematic} are quantitatively relevant only on an intermediate interval of $s_k$.

We now translate curvature patterns into allocation implications.

\begin{proposition}[Minority Voter Allocation and Welfare]\label{proposition-total-welfare-voter-distribution}
Fix total minority population $N_{mD}$ and district sizes $N_k$. Let minority welfare curvature satisfy
\begin{equation}
\frac{\partial^2 W(D)}{\partial s_k^2} = C_k(s_k) + I_k(s_k).
\end{equation}
\begin{enumerate}
\item Global Convexity. If $\frac{\partial^2 W(D)}{\partial s_k^2}>0$ for all feasible $s_k$, then minority welfare is globally convex in concentration and optimal districting concentrates minority voters in as few districts as feasible.

\item Global Concavity. If $\frac{\partial^2 W(D)}{\partial s_k^2}<0$ for all feasible $s_k$, then minority welfare is globally concave in concentration and optimal districting disperses minority voters more evenly across districts.

\item Nonmonotonicity. If $\frac{\partial^2 W(D)}{\partial s_k^2}$ changes sign over feasible $s_k$, then minority welfare is nonmonotonic in concentration; curvature reversal can be driven by
    \begin{enumerate}
    \item opposing curvature contributions within the fixed-matchup benchmark $C_k(s_k)$, and/or
    \item overturning general-equilibrium feedback $I_k(s_k)$ in electorally sensitive \emph{tipping} districts (Lemma~\ref{lemma_tipping_region}).
    \end{enumerate}
\end{enumerate}

\end{proposition}

Our equilibrium results show that optimal redistricting is governed by a simple structure: total welfare curvature equals a benchmark term that reflects descriptive and substantive representation under fixed matchups, plus an endogenous feedback term. When the benchmark curvature $C_k$ and the feedback $I_k$ share the same sign, or when the feedback is negligible, the isolated-channel logic of packing or cracking applies. When the curvature contributions of descriptive and substantive representation diverge or when feedback overturns curvature in competitive districts, minority welfare is nonmonotonic in minority concentration. Hence, the welfare consequences of packing and cracking depend not only on minority leverage and preference weights, but also on the primary system and on how district composition reshapes the electoral environment that determines political influence. 

Table~\ref{tbl:global-curvature-distribution} summarizes our equilibrium results from above, highlighting nonmonotonocities in minority welfare driven by the isolated channels, equilibrium feedback, or both, and their implications for optimal minority voter allocation.

\begin{landscape}
\begin{table}
\footnotesize
\begin{tabular}{lcccc}
\toprule
\textbf{Cases}           & \textbf{Competition Channel}     & \textbf{Selection Channel}               & \textbf{Benchmark}                                   & \textbf{Implications}                                                         \\ \midrule
\multicolumn{5}{l}{\textit{Fixed-matchup benchmark (isolated channels):} $W_{ss}=C_k=W^s_{ss}+W^{c,0}_{ss}$}     \\
Aligned Convex                & $W^{c,0}_{ss}>0$  & $W^s_{ss}>0$  & $C_k>0$    & Packing (concentration)  \\
Aligned Concave               & $W^{c,0}_{ss}<0$ & $W^s_{ss}<0$  & $C_k<0$     & Cracking (dispersion)     \\
Divergence I                   & $W^{c,0}_{ss}>0$                 & $W^s_{ss}<0$ & $C_k$ ambiguous and & Trade-off and \\
Divergence II                  & $W^{c,0}_{ss}<0$                 & $W^s_{ss}>0$  & may be single-crossing   & may be nonmonotonic    \\\midrule
\multicolumn{5}{l}{\textit{General equilibrium feedback: $W_{ss}=C_k+I_k$ with Lemma \ref{lemma_tipping_region} identifying $I_k$ first-order or second-order effects}}  \\
Safe districts                 & \multicolumn{2}{c}{Four cases above} & $|I_k|$ second-order  & Benchmark preserved \\
Tipping districts: reinforcing & \multicolumn{2}{c}{Four cases above} & $\text{sign}(I_k)=\text{sign}(C_k)$  & Amplified packing/cracking \\
Tipping districts: reversing & \multicolumn{2}{c}{Four cases above} & $\sign(I_k)\neq\sign(C_k)$, $|I_k|>|C_k|$ locally  & May be nonmonotonic \\ \bottomrule
\end{tabular}
\caption{General Equilibrium Curvature and Minority Voter Distribution.}
\label{tbl:global-curvature-distribution}
\end{table}
\end{landscape}

\section{Conclusion and Policy Implications}\label{section_conclusion}

Electoral institutions shape not only who wins office but also how policy benefits are distributed across voter groups. Building on the distributive politics tradition \citep{lindbeckweibull1987,myerson1993, dixitlondregan1995, dixitlondregan1996}, which emphasizes how electoral competition determines the allocation of targeted benefits, we show how districting itself becomes part of that distributive mechanism by reshaping voter responsiveness and political leverage across voter groups. District design, therefore, affects electoral outcomes and the equilibrium distribution of policy benefits. Redistricting provides a particularly important example of this institutional design problem, with primaries shaping how candidate competition translates district composition into political leverage. 

Our analysis highlights three empirical and policy-relevant implications of designing minority-majority districts, as well as the potential trade-offs between descriptive and substantive representation. First, the welfare effects of concentrating or dispersing minority voters depend on electoral competitiveness. Identical districting strategies may increase minority welfare in competitive districts but reduce it in safe districts where electoral responsiveness is weaker. Second, minorities' valuation of descriptive versus partisan representation ($\beta$) interacts with the primary system, shaping whether concentration generates increasing or decreasing returns. Third, the distribution of voter-group power (captured by $\kappa_i$, $\phi_i(.)$, and $\epsilon$) shapes whether minorities benefit from greater distributional leverage in a few districts or in many.

Our framework also provides testable predictions. In particular, the welfare effects of minority concentration should vary systematically with electoral competitiveness ($\mu_i^e$ and $\Phi_i^e$), with descriptive versus partisan voting behavior, with crossover voting ($\Delta C_k$), and with the responsiveness of voter groups to distributive incentives.

In this sense, the model clarifies the puzzle motivating the paper: districting that increases descriptive minority representation can simultaneously weaken substantive minority representation by altering voter responsiveness and the political leverage of different voter groups. More broadly, our analysis shows that electoral institutions, such as districting, not only shape who gets elected but also determine how political competition translates into the equilibrium distribution of policy benefits.

\newpage

\begingroup
		\setlength{\bibsep}{10pt}
    \setstretch{.75}
 \bibliographystyle{ecta}
\bibliography{References_Redistricting-Representation}
\endgroup


\newpage
\appendix

\newpage
\section{Appendix: Derivations and Proofs}\label{section_appendix_proofs}

\subsection{Proof of Lemma~\ref{lemma-platforms}}\label{appendix_proof-lemma-platforms}

The derivation and solution of the subgame described in Section \ref{subsection_order-of-play} follow \citet{dixitlondregan1996} and incorporate the two-stage electoral process from our model. First, we derive candidates' platforms, describing the voter group and the individual benefits; second, we compare these benefits across different primary structures; lastly, we describe the platforms of the general elections, illustrating distributive benefits for voters. 

\paragraph{Elections and Voter Groups} A minority Democrat and a nonminority Democrat face off in a primary, and then the winner and a Republican face each other in the general election. Candidates adopt platforms to maximize their votes, subject to an allocation of district benefits across voter groups. The voter groups in a closed primary are $\Theta \in \{mD, nD\}$, in an open primary $\Theta \in \{mD, nD, R\}$, and in a general election $\Theta \in \{mD, nD, R\}$.

\paragraph{Candidate Platforms} Two candidates announce simultaneously in each election. Candidate $1$'s problem is
\begin{equation}\label{app_problem-candidate1}
    \max V_1^e(k) = \sum_{i\in\Theta} N_{i,k}\Phi_i(x_i^e(k)) \text{ s.t. } \sum_{i\in\Theta}T_{i,1,k}B_k = \sum_{i\in\Theta}N_{i,k}b_{i,1,k} \leq B_k
\end{equation}
and candidate $2$'s is equivalently
\begin{equation}\label{app_problem-candidate2}
    \max V_2^e(k) = \sum_{i\in\Theta} N_{i,k}[1-\Phi_i(x_i^e(k))] \text{ s.t. } \sum_{i\in\Theta}T_{i,2,k}B_k = \sum_{i\in\Theta}N_{i,k}b_{i,2,k} \leq B_k.
\end{equation}

For the existence of a Nash equilibrium, Glicksberg's Theorem requires that each candidate's payoffs are a quasi-concave function of their strategy and a continuous function of other players' strategies. First, the distributive benefits for voters, $b_{ij}$, are an increasing linear function of the candidates' platforms, $T_{i,j,k}$. Second, the voters' cutoff for differences in candidates' promised benefits, $x_i^e(k)=\mu_i^e+\kappa_i\frac{b_{i,1,k}^{1-\epsilon}-b_{i,2,k}^{1-\epsilon}}{1-\epsilon}$, is increasing and concave in $b_{i,j,k}$. Finally, the expected numbers of votes, $V_1^e(k)=\sum_{i\in\Theta} N_{i,k}\Phi_i(x_i^e(k))$ and $V_2^e(k)=\sum_{i\in\Theta} N_{i,k}(1-\Phi_i(x_i^e(k)))$, are increasing in the cutoff $x_i^e(k)$ and concave due to the concavity of $\Phi_i(.)$. Hence, existence holds.

For the solution of the Nash equilibrium, we use Lagrange parameters $\lambda_1$ and $\lambda_2$ for each respective candidate and solve all first-order conditions simultaneously. Consider candidate $1$ first:
\begin{equation}
    L= \sum_{i\in\Theta} N_{i,k}\Phi_i(x_i^e(k)) + \lambda_1\left(B_k- \sum_{i\in\Theta}N_{i,k}b_{i,1,k}\right)
\end{equation}
with 
\begin{equation}
\frac{\partial L}{\partial T_{i,1,k}} = N_{i,k}\phi_i(x_i^e(k))u_i'(b_{i,1,k})\frac{\partial b_{i,1,k}}{\partial T_{i,1,k}} - \lambda_1N_{i,k}\frac{\partial b_{i,1,k}}{\partial T_{i,1,k}} = 0, 
\end{equation}
which can be written as 
\begin{equation}
    \lambda_1 = \phi_i(x_i^e(k))u_i'(b_{i,1,k}) \text{ } \Leftrightarrow \text{ } b_{i,1,k} = H_i\left(\frac{\lambda_1}{\phi_i(x_i^e(k))} \right),
\end{equation}
where $H_i(.)=(u_i')^{-1}(.)$ is the inverse of the marginal utility function. Since $u_i(.)$ is a decreasing function, $H_i(.)$ is decreasing as well, and there is a unique solution for $\lambda_1$. 

Candidate 2's problem is symmetric, and we have
\begin{equation}
    L= \sum_{i\in\Theta} N_{i,k}(1-\Phi_i(x_i^e(k))) + \lambda_2\left(B_k- \sum_{i\in\Theta}N_{i,k}b_{i,2,k}\right)
\end{equation}
with 
\begin{equation}
\frac{\partial L}{\partial T_{i,2,k}} = -N_{i,k}\phi_i(x_i^e(k))(-u_i'(b_{i,2,k}))\frac{\partial b_{i,2,k}}{\partial T_{i,2,k}} - \lambda_2N_{i,k}\frac{\partial b_{i,2,k}}{\partial T_{i,2,k}} = 0, 
\end{equation}
which can be written as 
\begin{equation}
    \lambda_2 = \phi_i(x_i^e(k))u_i'(b_{i,2,k}) \text{ } \Leftrightarrow \text{ } b_{i,2,k} = H_i\left(\frac{\lambda_2}{\phi_i(x_i^e(k))} \right),
\end{equation}
which provides a unique solution for $\lambda_2$.

The Lagrangian parameters are independent of candidates' characteristics, implying that both candidates face the same shadow value in equilibrium, $\lambda_1=\lambda_2$. As a result, a voter's marginal utility in distributive benefits is equal across both candidates: 
\begin{equation}
    \lambda_1=\lambda_2 \text{ } \Leftrightarrow \text{ } u_i'(b_{i,1,k}) = u_i'(b_{i,2,k}). 
\end{equation}
Due to $u_i(.)$ being a continuous, increasing function, we have that the distributive benefits are identical across both candidates, $b_{i,1,k}=b_{i,2,k}$, which implies that both candidates choose identical platforms, $T_{i,1,k}=T_{i,2,k}$, and distributive promises cancel each other out such that voters choose based on ideological alignments.

\begin{enumerate} 
\item \textbf{Group and Member Benefits} 
Now, we describe the distribution of district benefits across groups and their members. We use the first-order conditions above with the voters' utility function described by (\ref{voter-utility}) and (\ref{eq:utility_consumption_FOC-SOC}):
\begin{equation}
    \lambda_j = \phi_i(x_i^e(k))u_i'(b_{i,j,k}) \text{ } \Rightarrow \text{ } b^e_{i,j,k} = \left(\frac{\phi_i(x_i^e(k))\kappa_i}{\lambda_j} \right)^{\frac{1}{\epsilon}} = \frac{\pi_i^e}{\lambda_j^{1/\epsilon}} \text{ } \Leftrightarrow \text{ } \lambda_j^{1/\epsilon} = \frac{\pi_i^e}{b_{i,j,k}}.
\end{equation}
Applying $\lambda_1=\lambda_2$ and $b_{i,1,k}=b_{i,2,k}$, we get for group $i$ compared to group $h\neq i$ that
\begin{equation}\label{app-benefits-across_groups}
    b^e_{i,j,k}=\frac{\pi_i^e b_{h,j,k}}{\pi^e_h}.
\end{equation}
Using the budget constraint of $\sum_iN_{i,k}b_{i,j,k}=B_k$ with (\ref{app-benefits-across_groups}), we get
\begin{equation}\label{app-indiv-benefits}
b^e_{h,j,k} = \frac{\pi^e_h}{\sum_iN_{i,k}\pi^e_i}B_k,
\end{equation}
which provides the individual benefits for a member of group $i$ in (\ref{eq:individual-group-equilibrium-benefits}). The group shares follow from rearranging $b_{i,j,k} = (T_{i,j,k}*B_k)/N_{i,k}$ and applying (\ref{app-indiv-benefits}):
\begin{equation}
    T^e_{h,j,k} = \frac{b^e_{h,j,k}N_{h,k}}{B_k}=\frac{\pi^e_h N_{h,k}}{\sum_iN_{i,k}\pi_i^e},
\end{equation}
which completes the first statement with (\ref{eq:individual-group-equilibrium-benefits}) and (\ref{eq:group-power}) with $B_k=1$.

\item \textbf{Changing Platforms in Districts with Closed Primary} First, suppose a district has a closed primary with only Democratic voters, voters are myopic, and candidates can adjust platforms from the primary to the general election. The offered distributive benefits with an indicator for the electoral stage, $e=1$, are then 
\begin{eqnarray}
b^1_{mD,mD,k} &=& \frac{\pi^1_{mD}}{N_{mD,k}\pi^1_{mD}+N_{nD,k}\pi^1_{nD}}=b^1_{mD,nD,k}, \label{proof-distr-ben-mD0}\\
b^1_{nD,mD,k} &=& \frac{\pi^1_{nD}}{N_{mD,k}\pi^1_{mD}+N_{nD,k}\pi^1_{nD}}=b^1_{nD,nD,k}\label{proof-distr-ben-nD0}.
\end{eqnarray}

Second, suppose the winner of the closed primary, $j'$, faces a Republican candidate in the general election with Democratic and Republican voters. The offered distributive benefits for $e=2$ are then  
\begin{eqnarray}
b^2_{mD,j',k} &=& \frac{\pi^2_{mD}}{N_{mD,k}\pi^2_{mD}+N_{nD,k}\pi^2_{nD}+N_{R,k}\pi^2_{R}}=b^2_{mD,R,k},\label{proof-distr-ben-mD} \\
b^2_{nD,j',k} &=& \frac{\pi^2_{nD}}{N_{mD,k}\pi_{mD}^2+N_{nD,k}\pi^2_{nD}+N_{R,k}\pi^2_{R}}=b^2_{nD,R,k}, \label{proof-distr-ben-nD}\\
b^2_{R,j',k} &=& \frac{\pi^2_{R}}{N_{mD,k}\pi^2_{mD}+N_{nD,k}\pi^2_{nD}+N_{R,k}\pi^2_{R}}=b^2_{R,R,k}\label{proof-distr-ben-R}.
\end{eqnarray}

Comparing the offered benefits, we can state that $b^1_{mD,j',k}\neq b^2_{mD,j',k}$ and $b^1_{nD,j',k}\neq b^2_{nD,j',k}$, which completes the second statement of the lemma.

\item \textbf{General Election Platforms} 

The distributive benefits offered by unconstrained candidates (no reputation or commitment constraints) in a closed primary with myopic voters differ but converge to (\ref{proof-distr-ben-mD}), (\ref{proof-distr-ben-nD}), and (\ref{proof-distr-ben-R}) as shown above.

Next, suppose primary voters are forward-looking and anticipate the promises made in the general election when they cast their ballots in a closed primary. The anticipated benefits are then
\begin{eqnarray}
\tilde{b}^1_{mD,mD,k} &=& \frac{\pi^1_{mD}}{N_{mD,k}\pi^1_{mD}+N_{nD,k}\pi^1_{nD}+N_{R,k}\pi^1_{R}}=\tilde{b}^1_{mD,nD,k} \label{proof-distr-ben-mDrat}\\
\tilde{b}^1_{nD,mD,k} &=& \frac{\pi^1_{nD}}{N_{mD,k}\pi^1_{mD}+N_{nD,k}\pi^1_{nD}}=\tilde{b}^1_{nD,nD,k}\label{proof-distr-ben-nDrat},
\end{eqnarray}
which differ from (\ref{proof-distr-ben-mD0}) and (\ref{proof-distr-ben-nD0}), but are identical to (\ref{proof-distr-ben-mD}) and (\ref{proof-distr-ben-nD}).

Finally, suppose the primary is open to all three voter groups. The offered distributive benefits in the open primary are 
\begin{eqnarray}
\hat{b}^1_{mD,mD,k} &=& \frac{\pi^1_{mD}}{N_{mD,k}\pi^1_{mD}+N_{nD,k}\pi^1_{nD}+N_{R,k}\pi^1_{R}}=\hat{b}^1_{mD,nD,k},\label{proof-distr-ben-mD2} \\
\hat{b}^1_{nD,mD,k} &=& \frac{\pi^1_{nD}}{N_{mD,k}\pi^1_{mD}+N_{nD,k}\pi^1_{nD}+N_{R,k}\pi^1_{R}}=\hat{b}^1_{nD,nD,k}, \label{proof-distr-ben-nD2}\\
\hat{b}^1_{R,mD,k} &=& \frac{\pi^1_{R}}{N_{mD,k}\pi^1_{mD}+N_{nD,k}\pi^1_{nD}+N_{R,k}\pi^1_{R}}=\hat{b}^1_{R,nD,k},\label{proof-distr-ben-R2}
\end{eqnarray}
which are identical to (\ref{proof-distr-ben-mD}), (\ref{proof-distr-ben-nD}), and (\ref{proof-distr-ben-R}).

These three parts complete the third statement of the lemma.
 
\end{enumerate}

\subsection{Proof of Proposition~\ref{proposition-distributive-district-composition}}\label{proof_proposition-distributive-district-composition}

\begin{figure}[ht!]
\centering
    \subfigure[$\pi_{nD}<\pi_R$ and $\Pi_1\leq\Pi_2$.]{\label{app-fig:prop1Rpower}\includegraphics[scale=.6]{Figures/Simplex-Proposition1_Rpower.jpg}}
    \subfigure[$\pi_{nD}>\pi_R$ and $\Pi_1\leq\Pi_2$.]{\label{app-fig:prop1Dpower}\includegraphics[scale=.6]{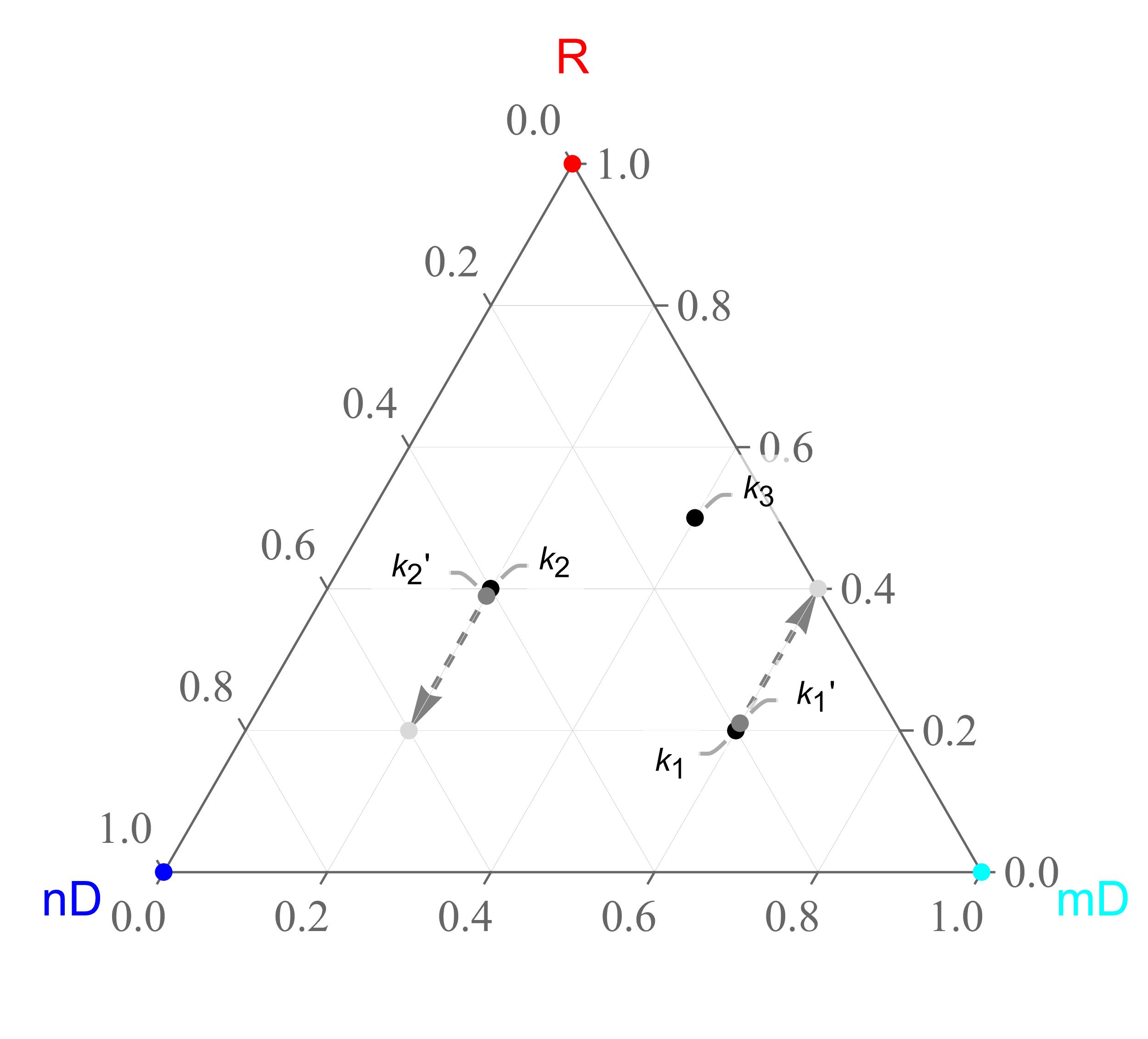}}
    \caption{Optimal Districting Process with $K=3$.}
\end{figure}

Consider a valid districting scheme $\mathbf{D}$, and assume to the contrary that $N_{i, k} > 0$ for all $i\in\Theta$ and $k$. First, assume that Republicans are more powerful than nonminority Democrats, $\pi_{nD}<\pi_R$. Consider two districts $k_1$ and $k_2$, with $N_{mD,1}$ minority voters in $k_1$ and $N_{mD,2}$ in $k_2$, and assume that $N_{mD,1}>N_{mD,2}$. Because district populations are large and voter shares are continuous, we evaluate marginal reallocations of minority voters across districts using local comparative statics. This is illustrated in Figure~\ref{app-fig:prop1Rpower}. 

We now move one Republican voter from $k_1$ to $k_2$ and one nonminority-Democrat voter from $k_2$ to $k_1$, while holding minority voters constant in each district. Such a change preserves the validity of the districting scheme, and the arrows in Figure~\ref{app-fig:prop1Rpower} illustrate the direction of changes. For the initial valid districting scheme to be optimal, no such reallocation can increase minority voters' distributive benefits. Hence, we compare any gains and losses across the two districts by considering marginal changes in district composition since district populations are large and we are interested in the direction of profitable local reallocations.

Considering the changes in minority benefits from (\ref{eq:foc-minority-benefits-nD-voters}), minority voters' distributive benefits in $k_1$ are increasing when a Republican is replaced with a less powerful nonminority-Democratic voter, by
\begin{equation}\label{eq:appendix-k1gain}
 N_{mD,1}\cdot\frac{\partial f}{\partial N_{nD,1}} = N_{mD,1}^2 \cdot \frac{\pi_{mD}(\pi_R-\pi_{nD})}{\Pi_1^2};
\end{equation}
while minority voters' distributive benefits in $k_2$ are decreasing when a more powerful Republican replaces the nonminority-Democrat voter, by
\begin{equation}\label{eq:appendix-k2loss}
 N_{mD,2}\frac{\partial f}{\partial N_{nD,2}} = N_{mD,2}^2\cdot\frac{\pi_{mD}(\pi_R-\pi_{nD})}{\Pi_2^2}.
\end{equation}
Comparing (\ref{eq:appendix-k1gain}) with (\ref{eq:appendix-k2loss}), we get
\begin{equation}
\frac{N_{mD,1}^2}{\Pi_1^2} \gtreqless \frac{N_{mD,2}^2}{\Pi_2^2}. 
\end{equation}
Given the assumption of $N_{mD,1}>N_{mD,2}$ and focusing on small changes in population shares that do not alter the sign of the net payoff change significantly, the minority gains in $k_1$ outweigh the minority losses in $k_2$ if $\Pi_1^2\leq \Pi_2^2$, and such redistricting would be optimal as it increases net gains for minority voters. 

Second, assume that nonminority-Democrat voters are more powerful than Republican voters, $\pi_{nD}>\pi_{R}$, and the number of minorities differs, $N_{mD,1}>N_{mD,2}$. Then it would be beneficial to move a nonminority Democrat from $k_1$ to $k_2$, and a Republican from $k_2$ to $k_1$, if the minority gains in $k_1$ are greater than the minority losses in $k_2$. This comparison follows again
\begin{equation}
\frac{N_{mD,1}^2}{\Pi_1^2} \gtreqless \frac{N_{mD,2}^2}{\Pi_2^2} \text{ with } \pi_{nD}>\pi_{R}. 
\end{equation}
If the minority-concentrated district is less powerful, then this would be beneficial. Figure~\ref{app-fig:prop1Dpower} illustrates this process.

Hence, the proposed districting scheme $\mathbf{D}$ cannot be optimal as any alternative scheme with two interior districts offers profitable swaps. By reiterating the process, at most one district lies in the interior of the simplex $S^2$. Furthermore, as a district's power decreases when less powerful nonminority voters replace more powerful nonminority voters, $\frac{\partial \Pi_k}{\partial N_{nD}}=\pi_{nD}-\pi_R$, minority populated districts with low district power or nonminority populated districts with high district power will not lie in the interior of $S^2$.

\subsection{Hessian Matrix for Distributive Benefits}\label{details_hessian-distributive-benefits}

We calculate the determinants of the principal minors of the Hessian matrix: 
\begin{equation}
H = \begin{bmatrix}
    \frac{\partial^2 f}{\partial N_{mD,k}^2} & \frac{\partial^2 f}{\partial N_{mD,k}N_{nD,k}} \\
    \frac{\partial^2 f}{\partial N_{nD,k}N_{mD,k}} & \frac{\partial^2 f}{\partial N_{nD,k}^2}
  \end{bmatrix}.
\end{equation}
with 
\begin{eqnarray}
\frac{\partial^2 f}{\partial N_{mD,k}^2} &=& \frac{2\pi_{mD}(\pi_R-\pi_{mD})\left(\pi_RN_k+(\pi_{nD}-\pi_R)N_{nD,k}\right)}{\Pi_k^3}, \\
\frac{\partial^2 f}{\partial N_{mD,k}N_{nD,k}} &=& \frac{\pi_{mD}(\pi_{nD}-\pi_R)\left((\pi_{mD}-\pi_R)N_{mD,k} + (\pi_R-\pi_{nD})N_{nD,k} - \pi_RN_k \right)}{\Pi_k^3}, \\
\frac{\partial^2 f}{\partial N_{nD,k}^2} &=&  \frac{2\pi_{mD}(\pi_{nD}-\pi_R)^2N_{mD,k}}{\Pi_k^3},
\end{eqnarray}
and
\begin{equation}
det(H) = \frac{\pi_{mD}^2(\pi_R-\pi_{nD})^2}{\Pi_k^4}.
\end{equation}

The determinant of the entire $H$ matrix is positive on the interior of $S^2$ for $\pi_{nD}\neq\pi_{R}$, but the value of $\frac{\partial^2 f}{\partial N_{mD,k}^2}$ is indeterminate, indicating that the $H$ matrix can be positive definite, negative definite, or neither, depending on the parameter values. For optimization, the surface could be either concave or convex. 


\subsection{Proof of Lemma \ref{lemma-distributive-districting-payoff-simplex}}\label{proof_lemma-distributive-districting-payoff-simplex}

Consider a valid districting scheme $\mathbf{D}$ and a district $\mathbf{d^\ast}=(N_{mD},N_{nD},N_{R})$. Define $s_k=N_{mD,k}/N_k$ and $t_k=N_{nD,k}/(N_{nD,}+N_{R,k})$, and let $l=\{\mathbf{d} \in \mathcal{D}|t=N_{nD}/(N_{nD}+N_R)\}$. Here, $l$ is a line running through $\mathbf{d^\ast}$, connecting it to $(1,0,0)$, which is the corner of $S^2$ where the district comprises minority voters only (``mD'' in triangles) while keeping the ratio of nonminority voters constant throughout. Applying (\ref{eq:minority-benefits-population-power-orig}) divided by $N_k$ with $s_k=N_{mD,k}/N_k$, $(1-s_k)*t_k=\frac{N_{nD,k}+N_{R,k}}{N_k}*\frac{N_{nD,k}}{N_{nD,k}+N_{R,k}}=N_{nD,k}/N_k$, and $(1-s_k)*(1-t_k)=\frac{N_{nD,k}+N_{R,k}}{N_k}*\frac{N_{R,k}}{N_{nD,k}+N_{R,k}}=N_{R,k}/N_k$, the parameterized path for the minority payoff function is 
\begin{equation}
g(s_k)=\frac{\pi_{mD}s_k}{\pi_{mD}s_k+ \pi_{nD}(1-s_k)t_k + \pi_{R}(1-s_k)(1-t_k)}.
\end{equation}
This parameterization traces all feasible reallocations holding the ratio of nonminority voters fixed. Note that the denominator is positive, and we can evaluate the curvature of the path with its second derivative with respect to $t$:
\begin{equation}
g''(s_k)=-\frac{2\pi_{mD}\overbrace{\left(t_k\pi_{nD}+(1-t_k)\pi_R\right)}^{(+)}\overbrace{\left( \pi_{mD} - t_k\pi_{nD}-(1-t_k)\pi_R\right)}^{(?)}}{(\underbrace{\pi_{mD}s_k+ \pi_{nD}(1-s_k)t_k + \pi_{R}(1-s_k)(1-t_k)}_{(+)})^3}.
\end{equation}
The second derivative is negative if $\pi_{mD}>t_k\pi_{nD}+(1-t_k)\pi_R$ -- i.e., when minorities' power is greater than the weighted average of the other groups' powers, based on district population. Hence, $\pi_{mD}=\max_{i\in\Theta}\{\pi_i \}$ implies that $g''(s_k)<0$ for all $s_k\in(0,1)$, indicating that the entire surface is concave on the interior of $S^2$. Conversely, since $t_k\in(0,1)$ on the interior of $S^2$, the condition $\pi_{mD}=\min_{i\in\Theta}\{\pi_i\}$ implies $\pi_{mD}<t_k\pi_{nD}+(1-t_k)\pi_R$ and hence $g''(s_k)>0$ for all $s_k\in(0,1)$, and the surface is convex on the interior of $S^2$.

\subsection{Proof of Proposition \ref{proposition_distributive-districting-distribution}}\label{proof_proposition_distributive-districting-distribution}

Let $s_k\equiv N_{mD,k}/N_k$ with $N_k$ identical across all districts and $\sum_k N_k s_k=N_{mD}$ be fixed. Under the competition channel with fixed matchups, minority distributive welfare can be rewritten as
\begin{equation}
W^{c}(D)= \sum_{k=1}^Ks_kN_k\cdot\kappa_{mD}\cdot\frac{b_{mD,k}^{1-\epsilon}}{1-\epsilon} = \sum_{k=1}^K N_k\cdot w^c(s_k), \quad w^c(s_k)\equiv s_k\cdot\kappa_{mD}\cdot\frac{b_{mD,k}^{1-\epsilon}}{1-\epsilon}.
\end{equation}
Lemma~\ref{lemma-distributive-districting-payoff-simplex} implies that $w^c(s_k)$ is convex when $\pi_{mD}=\min_{i\in\Theta}\{\pi_i\}$ and concave when $\pi_{mD}=\max_{i\in\Theta}\{\pi_i\}$. 

\paragraph{1) Minority Power Lowest}

Let $s_{k_1}>s_{k_2}$ and consider a mean-preserving transfer $\delta>0$ from district $k_2$ to $k_1$ (for $\delta$ small enough).

If $w^c(s_k)$ is convex, then $w^c(s_{k_1}+\delta) + w^c(s_{k_2}-\delta) \geq w^c(s_{k_1}) + w^c(s_{k_2})$, so concentrating minority voters weakly increases $W^c(D)$. By repeating the process, minority welfare is maximized at boundary allocations -- i.e., by concentrating minority voters as much as feasible.

\paragraph{2) Minority Power Highest}

Let $s_{k_1}>s_{k_2}$ and consider a mean-preserving transfer $\delta>0$ from district $k_1$ to $k_2$ (again for small enough $\delta$).

If $w^c(s_k)$ is concave, then $w^c(s_{k_1}-\delta) + w^c(s_{k_2}+\delta) \geq w^c(s_{k_1}) + w^c(s_{k_2})$, so dispersing minority voters weakly increases $W^c(D)$. By repeating the process, minority welfare is maximized under more equal allocations, i.e., by dispersing minority voters as evenly as possible.

\subsection{Proof of Proposition \ref{proposition_min-candidate-selection}}\label{proof_proposition_min-candidate-selection}

Isolating the selection channel, we write the winning probabilities for a minority candidate:
\begin{eqnarray}
\text{Closed Primary: } \Psi^1_{mD,k}&=&\frac{\Phi^1_{mD}(\mu_{mD}^1)N_{mD,k}+\Phi^1_{nD}(\mu_{nD}^1)N_{nD,k}}{N_{mD,k}+N_{nD,k}}; \\
\text{Open Primary: } \tilde{\Psi}^1_{mD,k}&=& \frac{\Phi^1_{mD}(\mu_{mD}^1)N_{mD,k}+\Phi^1_{nD}(\mu_{nD}^1)N_{nD,k}+\Phi^1_R(\mu_R^1)N_{R,k}}{N_{mD,k}+N_{nD,k}+N_{R,k}};\\
\text{General: } \Psi^2_{mD,k}&=&\frac{\Phi^2_{mD}(\mu_{mD}^2)N_{mD,k}+\Phi^2_{nD}(\mu_{nD}^2)N_{nD,k}+\Phi^2_R(\mu_R^2)N_{R,k}}{N_{mD,k}+N_{nD,k}+N_{R,k}};\\
\text{District Win: } \Psi_{mD,k}&=& \Psi^1_{mD,k} \cdot \Psi^2_{mD,k}  \\
\text{ or } \tilde{\Psi}_{mD,k}&=& \tilde{\Psi}^1_{mD,k} \cdot \Psi^2_{mD,k}.
\end{eqnarray}

Using these probabilities, we derive the following four statements.

\paragraph{1) Minority Population} First, we focus on the local effects of increasing the number of minority voters in a given district:
\begin{itemize}
 \item Closed primaries
\begin{eqnarray}
\frac{\partial \Psi_{mD,k}}{\partial N_{mD,k}} &=& \overbrace{\frac{\partial \Psi^1_{mD,k}}{\partial N_{mD,k}}\cdot\Psi^2_{mD,k}}^{primary} + \overbrace{\Psi^1_{mD,k}\cdot\frac{\partial \Psi^2_{mD,k}}{\partial N_{mD,k}}}^{general} \\
=&& \underbrace{\frac{(\Phi^1_{mD}(\mu_{mD}^1)-\Phi^1_{nD}(\mu_{nD}^1))N_{nD,k}}{(N_{mD,k}+N_{nD,k})^2}\cdot\underbrace{\Psi^2_{mD,k}}_{(\geq 0)}}_{primary: (+)} \nonumber \\
&&+ \underbrace{\underbrace{\Psi^1_{mD,k}}_{(\geq 0)}\cdot\frac{(\Phi^2_{mD}(\mu_{mD}^G)-\Phi^2_{nD}(\mu_{nD}^G))N_{nD,k}+(\Phi^2_{mD}(\mu_{mD}^G)-\Phi^2_{R}(\mu_{R}^G))N_{R,k}}{(N_{mD,k}+N_{nD,k}+N_{R,k})^2}}_{general: (+)}\geq 0.
\end{eqnarray}
The assigned signs follow immediately from applying \eqref{eq:min-candidate-order}.

\item Open primaries
\begin{eqnarray}
\frac{\partial \tilde{\Psi}_{mD,k}}{\partial N_{mD,k}} &=& \overbrace{\frac{\partial \tilde{\Psi}^1_{mD,k}}{\partial N_{mD,k}}\cdot\Psi^2_{mD,k}}^{primary} + \overbrace{\tilde{\Psi}^1_{mD,k}\cdot\frac{\partial \Psi^2_{mD,k}}{\partial N_{mD,k}}}^{general} \\
&=& \underbrace{\frac{(\Phi^1_{mD}(\mu_{mD}^1)-\Phi^1_{nD}(\mu_{nD}^1))N_{nD,k}+(\Phi^1_{mD}(\mu_{mD}^1)-\Phi^1_{R}(\mu_{R}^1))N_{R,k}}{(N_{mD,k}+N_{nD,k}+N_{R,k})^2}\cdot\underbrace{\Psi^2_{mD,k}}_{(\geq 0)}}_{primary: (+)} \nonumber \\
&&+ \underbrace{\underbrace{\tilde{\Psi}^1_{mD,k}}_{(\geq 0)}\cdot\frac{(\Phi^2_{mD}(\mu_{mD}^2)-\Phi^2_{nD}(\mu_{nD}^2))N_{nD,k}+(\Phi^2_{mD}(\mu_{mD}^2)-\Phi^2_{R}(\mu_{R}^2))N_{R,k}}{(N_{mD,k}+N_{nD,k}+N_{R,k})^2}}_{general: (+)}\geq 0.
\end{eqnarray}

Both illustrate the weakly positive effect of the number of minority voters on the likelihood of minority candidate success.

\end{itemize}

\paragraph{2) Nonminority Democrats} Second, we focus on the local effects of increasing the number of nonminority Democratic voters in a given district:
\begin{itemize}
 \item Closed primaries
\begin{eqnarray}
\frac{\partial \Psi_{mD,k}}{\partial N_{nD,k}} &=& \overbrace{\frac{\partial \Psi^1_{mD,k}}{\partial N_{nD,k}}\cdot\Psi^2_{mD,k}}^{primary} + \overbrace{\Psi^1_{mD,k}\cdot\frac{\partial \Psi^2_{mD,k}}{\partial N_{nD,k}}}^{general} \\
&=& \underbrace{\frac{(\Phi^1_{nD}(\mu_{nD}^1)-\Phi^1_{mD}(\mu_{mD}^1))N_{mD,k}}{(N_{mD,k}+N_{nD,k})^2}\cdot\underbrace{\Psi^2_{mD,k}}_{(\geq 0)}}_{primary: (-)} \\
&&+ \underbrace{\underbrace{\Psi^1_{mD,k}}_{(\geq 0)}\cdot\frac{(\Phi^2_{nD}(\mu_{nD}^2)-\Phi^2_{mD}(\mu_{mD}^2))N_{mD,k}+(\Phi^2_{nD}(\mu_{nD}^2)-\Phi^2_{R}(\mu_{R}^2))N_{R,k}}{(N_{mD,k}+N_{nD,k}+N_{R,k})^2}}_{general: (+/-)} \gtreqless 0.
\end{eqnarray}

\item Open primaries
\begin{eqnarray}
\frac{\partial \tilde{\Psi}_{mD,k}}{\partial N_{nD,k}} &=& \overbrace{\frac{\partial \tilde{\Psi}^1_{mD,k}}{\partial N_{nD,k}}\cdot\Psi^2_{mD,k}}^{primary} + \overbrace{\tilde{\Psi}^1_{mD,k}\cdot\frac{\partial \Psi^2_{mD,k}}{\partial N_{nD,k}}}^{general} \\
&=& \underbrace{\frac{(\Phi^1_{nD}(\mu_{nD}^1)-\Phi^1_{mD}(\mu_{mD}^1))N_{mD,k}+(\Phi^1_{nD}(\mu_{nD}^1)-\Phi^1_{R}(\mu_{R}^1))N_{R,k}}{(N_{mD,k}+N_{nD,k}+N_{R,k})^2}\cdot\underbrace{\Psi^2_{mD,k}}_{(\geq 0)}}_{primary: (-)} \\
&&+ \underbrace{\underbrace{\tilde{\Psi}^1_{mD,k}}_{(\geq 0)}\cdot\frac{(\Phi^2_{nD}(\mu_{nD}^G)-\Phi^2_{mD}(\mu_{mD}^2))N_{mD,k}+(\Phi^2_{nD}(\mu_{nD}^2)-\Phi^2_{R}(\mu_{R}^2))N_{R,k}}{(N_{mD,k}+N_{nD,k}+N_{R,k})^2}}_{general: (+/-)} \gtreqless 0.
\end{eqnarray}

\end{itemize}

\paragraph{3) Nonminority Democrats Replacing Republicans -- Open Primaries} Third, we examine the local effects of replacing Republican voters with nonminority Democratic voters, holding $N_k$ and $N_{mD,k}$ fixed. We replace $N_{R,k}=N_k-N_{mD,k}-N_{nD,k}$ and take the derivative with respect to the number of nonminority-Democratic voters and get:
\begin{itemize}
\item Closed primaries
\begin{eqnarray}
\left.\frac{\partial \Psi_{mD,k}}{\partial N_{nD,k}}\right|_{N_k,N_{mD,k}} &=& \overbrace{\frac{\partial \Psi^1_{mD,k}}{\partial N_{nD,k}} \cdot\Psi^2_{mD,k}}^{primary} + \overbrace{\Psi^1_{mD,k}\cdot\frac{\partial \Psi^2_{mD,k}}{\partial N_{nD,k}}}^{general} \\
&=& \underbrace{\frac{(\Phi^1_{nD}(\mu_{nD}^1)-\Phi^1_{mD}(\mu_{mD}^1))N_{mD,k}}{(N_{mD,k}+N_{nD,k})^2}\cdot\underbrace{\Psi^2_{mD,k}}_{(\geq 0)}}_{primary: (-)} + \underbrace{\underbrace{\Psi^1_{mD,k}}_{(\geq 0)}\cdot\frac{\Phi^2_{nD}(\mu_{nD}^2)-\Phi^2_R(\mu_{R}^2)}{N_k}}_{general: (+)} \gtreqless 0,
\end{eqnarray}
where the minority candidate's chances decrease in the primary but increase in the general election.
\item Open primaries
\begin{eqnarray}
\left.\frac{\partial \tilde{\Psi}_{mD,k}}{\partial N_{nD,k}}\right|_{N_k,N_{mD,k}} &=& \overbrace{\frac{\partial \tilde{\Psi}^1_{mD,k}}{\partial N_{nD,k}} \cdot\Psi^2_{mD,k}}^{primary} + \overbrace{\tilde{\Psi}^1_{mD,k}\cdot\frac{\partial \Psi^2_{mD,k}}{\partial N_{nD,k}}}^{general} \\
&=& \underbrace{\frac{\Phi^1_{nD}(\mu_{nD}^1)-\Phi^1_{R}(\mu_{R}^1)}{N_k}\cdot\underbrace{\Psi^2_{mD,k}}_{(\geq 0)}}_{primary: (+)} + \underbrace{\underbrace{\tilde{\Psi}^1_{mD,k}}_{(\geq 0)}\cdot\frac{\Phi^2_{nD}(\mu_{nD}^2)-\Phi^2_{R}(\mu_{R}^2)}{N_k}}_{general: (+)} \geq 0,
\end{eqnarray}
where the minority candidate's chances increase in the primary and general elections when $nD$ voters are more supportive of the minority candidate than $R$ voters.

Hence, the weakly positive effect of replacing Republicans with nonminority Democrats holds under open primaries.
\end{itemize}

\paragraph{4) Two-Stage Election and Convexity} Let $s_k \equiv N_{mD,k}/N_k$ with $N_k$ fixed. Furthermore, define $t_k\equiv N_{nD,k}/(N_{nD,k}+N_{R,k})$ with $t_k\in(0,1)$ for the interior. We can write district population shares as 
\begin{eqnarray}
\text{Minority Democratic: } \frac{N_{mD,k}}{N_k}&=& s_k; \\
\text{Nonminority Democratic: } \frac{N_{nD,k}}{N_k} &=& (1-s_k)t_k; \\
\text{Republican: } \frac{N_{R,k}}{N_k} &=& (1-s_k)(1-t_k). 
\end{eqnarray}
Using these district-population shares, we can rewrite a minority candidate's winning probabilities
\begin{eqnarray}
\text{Closed Primary: } \Psi^1_{mD,k}&=&\frac{\Phi^1_{mD}(\mu_{mD}^1)s_k+\Phi^1_{nD}(\mu_{nD}^1)(1-s_k)t_k}{s_k+(1-s_k)t_k} \nonumber \\
&=&\frac{(\Phi^1_{mD}(\mu_{mD}^1)-\Phi^1_{nD}(\mu_{nD}^1)t_k)s_k+\Phi^1_{nD}(\mu_{nD}^1)t_k}{t_k+(1-t_k)s_k}; \\
\text{Open Primary: } \tilde{\Psi}^1_{mD,k}&=& \Phi^1_{mD}(\mu_{mD}^1)s_k+\Phi^1_{nD}(\mu_{nD}^1)(1-s_k)t_k+\Phi^1_R(\mu_R^1)(1-s_k)(1-t_k); \\
\text{General: } \Psi^2_{mD,k}&=&\Phi^2_{mD}(\mu_{mD}^2)s_k+\Phi^2_{nD}(\mu_{nD}^2)(1-s_k)t_k+\Phi^2_R(\mu_R^2)(1-s_k)(1-t_k). 
\end{eqnarray}

We can now evaluate the derivatives for convexity:
\begin{itemize}
\item Open primaries
\begin{eqnarray}
\frac{\partial^2 \tilde{\Psi}_{mD,k}}{\partial s_k^2} &=& \overbrace{\underbrace{\frac{\partial^2 \tilde{\Psi}^1_{mD,k}}{\partial s_k^2}}_{=0}\cdot\Psi^2_{mD,k}+\frac{\partial \tilde{\Psi}^1_{mD,k}}{\partial s_k}\cdot\frac{\partial \Psi_{mD,k}^2}{\partial s_k}}^{primary} + \overbrace{\frac{\partial \tilde{\Psi}^1_{mD,k}}{\partial s_k}\cdot\frac{\partial \Psi^2_{mD,k}}{\partial s_k}+\tilde{\Psi}^1_{mD,k}\cdot\underbrace{\frac{\partial^2 \Psi^2_{mD,k}}{\partial s_k^2}}_{=0}}^{general}\nonumber \\
&=& 2\cdot\frac{\partial \tilde{\Psi}^1_{mD,k}}{\partial s_k}\cdot\frac{\partial \Psi_{mD,k}^2}{\partial s_k} \nonumber \\
&=& 2\cdot\left(\underbrace{\Phi_{mD}^1(\mu_{mD}^1) - [t_k\Phi_{nD}^1(\mu_{nD}^1) + (1-t_k)\Phi_{R}^1(\mu_{R}^1)]}_{(+)}\right)\\
&&\cdot\left(\underbrace{\Phi_{mD}^2(\mu_{mD}^2) - [t_k\Phi_{nD}^2(\mu_{nD}^2) +(1-t_k)\Phi_{R}^2(\mu_{R}^2)]}_{(+)}\right)>0;
\end{eqnarray}
hence, $\tilde{\Psi}_{mD,k}$ is convex in minority population share $s_k$ on $S^2$.

\item Closed primaries
\begin{eqnarray}
\frac{\partial^2 \Psi_{mD,k}}{\partial s_k^2} &=& \overbrace{\frac{\partial^2 \Psi^1_{mD,k}}{\partial s_k^2}\cdot\Psi^2_{mD,k}+\frac{\partial \Psi^1_{mD,k}}{\partial s_k}\cdot\frac{\partial \Psi_{mD,k}^2}{\partial s_k}}^{primary} + \overbrace{\frac{\partial \Psi^1_{mD,k}}{\partial s_k}\cdot\frac{\partial \Psi^2_{mD,k}}{\partial s_k}+\Psi^1_{mD,k}\cdot\underbrace{\frac{\partial^2 \Psi^2_{mD,k}}{\partial s_k^2}}_{=0}}^{general}\nonumber \\
&=& \frac{\partial^2 \Psi^1_{mD,k}}{\partial s_k^2}\cdot\Psi^2_{mD,k}+2\cdot\frac{\partial \Psi^1_{mD,k}}{\partial s_k}\cdot\frac{\partial \Psi_{mD,k}^2}{\partial s_k} \nonumber \\
&=& -\frac{2(1-t_k)t_k(\Phi_{mD}^1(\mu_{mD}^1)-\Phi_{nD}^1(\mu_{nD}^1))}{(t_k+(1-t_k)s_k)^3}\cdot\Psi^2_{mD,k} \nonumber \\
&&+ 2\cdot\frac{t_k(\Phi_{mD}^1(\mu_{mD}^1)-\Phi_{nD}^1(\mu_{nD}^1))}{(t_k+(1-t_k)s_k)^2}\cdot\left(\Phi_{mD}^2(\mu_{mD}^2) - [t_k\Phi_{nD}^2(\mu_{nD}^2) +(1-t_k)\Phi_{R}^2(\mu_{R}^2)]\right) \nonumber \\
&=& \underbrace{\frac{2t_k(\Phi_{mD}^1(\mu_{mD}^1)-\Phi_{nD}^1(\mu_{nD}^1))}{(t_k+(1-t_k)s_k)^3}}_{(+)}\cdot\left((\Phi_{mD}^2(\mu_{mD}^2) -\Phi_{nD}^2(\mu_{nD}^2))t_k-(1-t_k)\Phi_{R}^2(\mu_{R}^2) \right), \nonumber
\end{eqnarray}
which is positive when 
\begin{equation}\label{eq:proof-convexity-closed-primary}
(\Phi_{mD}^2(\mu_{mD}^2) -\Phi_{nD}^2(\mu_{nD}^2))t_k-(1-t_k)\Phi_{R}^2(\mu_{R}^2)>0;
\end{equation}
hence, for $\Phi_{mD}^2(\mu_{mD}^2) > \Phi_{nD}^2(\mu_{nD}^2)+\frac{1-t_k}{t_k}\Phi_{R}^2(\mu_{R}^2)$, $\Psi_{mD,k}$ is convex in minority population share $s_k$ on $S^2$. The surface is concave if the opposite of \eqref{eq:proof-convexity-closed-primary} holds. Notably, the sign of $\frac{\partial^2 \Psi_{mD,k}}{\partial s_k^2}$ is independent of $s_k$ and depends only on the nonminority composition $t_k$.
\end{itemize}

\subsection{Proof of Lemma \ref{lemma_minority-ideology-benefits}}\label{app_lemma_minority-ideology-benefits}

We separate the proof into an analysis for a state with closed primaries and then repeat the steps for a state with open primaries.

\paragraph{Closed primaries.} Minority voters' expected ideological benefit in district $k$ is
\begin{eqnarray}
E[\mu_{mD}\mid k]
&=& \Psi^1_{mD,k}\Psi^2_{mD,k} + \Psi^1_{nD,k}\Psi^3_{nD,k}\,\beta \nonumber\\
&=& \Psi^1_{mD,k}\Psi^2_{mD,k} + (1-\Psi^1_{mD,k})\Psi^3_{nD,k}\,\beta , \nonumber
\end{eqnarray}
where $\Psi^1_{nD,k}=1-\Psi^1_{mD,k}$ under a two-candidate primary and $e=3$ indicates a general election of $nD$ against $R$. We indicate that $\Phi_i^3(\mu_i^3)$ is group $i$'s support for candidate $nD$ against candidate $R$ in that general election. Using $N_{R,k}=N_k-N_{mD,k}-N_{nD,k}$ and substituting the vote-share expressions yields
\begin{eqnarray}
E[\mu_{mD}\mid k]
&=&
\left(\frac{\Phi_{mD}^1(\mu_{mD}^1)N_{mD,k} + \Phi_{nD}^1(\mu_{nD}^1)N_{nD,k}}{N_{mD,k}+N_{nD,k}}\right) \nonumber\\
&&\cdot
\left(\frac{\Phi_{mD}^2(\mu_{mD}^2)N_{mD,k} + \Phi_{nD}^2(\mu_{nD}^2)N_{nD,k} + \Phi_R^2(\mu_R^2)(N_k-N_{mD,k}-N_{nD,k})}{N_k}\right) \nonumber\\
&&+\left(1-\frac{\Phi_{mD}^1(\mu_{mD}^1)N_{mD,k} + \Phi_{nD}^1(\mu_{nD}^1)N_{nD,k}}{N_{mD,k}+N_{nD,k}}\right) \nonumber \\
&&\cdot
\left(\frac{\Phi_{mD}^3(\mu_{mD}^3)N_{mD,k} + \Phi_{nD}^3(\mu_{nD}^3)N_{nD,k} + \Phi_R^3(\mu_R^3)(N_k-N_{mD,k}-N_{nD,k})}{N_k}\right)\beta. \nonumber
\end{eqnarray}

The second derivative with respect to $N_{mD,k}$ is
\begin{equation}\label{app_eq-exp-min-utility-2nd}
\frac{\partial^2 E[\mu_{mD}\mid k]}{\partial N_{mD,k}^2}
=
\frac{2\overbrace{\big(\Phi_{mD}^1(\mu_{mD}^1)-\Phi_{nD}^1(\mu_{nD}^1)\big)}^{(+)}N_{nD,k}}{N_k\,(N_{mD,k}+N_{nD,k})^3}\;\varphi,
\end{equation}
where
\begin{eqnarray}
\varphi
&\equiv&
\overbrace{\big(\Phi_{mD}^2(\mu_{mD}^2)-\Phi_{nD}^2(\mu_{nD}^2)\big)}^{(+)}N_{nD,k}
-\big(\Phi_R^2(\mu_R^2)-\Phi_R^3(\mu_R^3)\beta\big)N_k \nonumber\\
&&-\big(\Phi_{mD}^3(\mu_{mD}^3)-\Phi_{nD}^3(\mu_{nD}^3)\big)\beta\,N_{nD,k}.
\end{eqnarray}
The assigned signs follow from Assumption~\eqref{eq:min-candidate-order}. Since the second derivative is affine in $\beta$, solving $\partial^2 E[\mu_{mD}\mid k]/\partial N_{mD,k}^2=0$ yields the threshold
\begin{equation}
\bar{\beta}^C =
\frac{(\Phi_{mD}^2-\Phi_{nD}^2)N_{nD,k}-\Phi_R^2 N_k}
{(\Phi_{mD}^3-\Phi_{nD}^3)N_{nD,k}-\Phi_R^3 N_k}.
\end{equation}

If the denominator is positive, then 
\begin{equation}
\frac{\partial^2 E[\mu_{mD}\mid k]}{\partial N_{mD,k}^2} > 0 \quad\Longleftrightarrow\quad\beta < \bar{\beta}^C.
\end{equation}

If the denominator is negative, the inequality reverses and
\begin{equation}
\frac{\partial^2 E[\mu_{mD}\mid k]}{\partial N_{mD,k}^2} > 0 \quad\Longleftrightarrow\quad \beta > \bar{\beta}^C.
\end{equation}

\paragraph{Open Primaries} Following the analysis for closed primaries, we adopt the same strategy here. The expected benefit for minority voters is
\begin{eqnarray}
E[\mu_{mD}\mid k] &=& \tilde{\Psi}_{mD,k}^1 \Psi_{mD,k}^2 + \tilde{\Psi}_{nD,k}^1\Psi_{nD,k}^3\beta = \tilde{\Psi}_{mD,k}^1 \Psi_{mD,k}^2 + (1-\tilde{\Psi}_{mD,k}^1)\Psi_{nD,k}^3\beta \nonumber \\
 &=& \left(\frac{\Phi_{mD}^1(\mu_{mD}^1)N_{mD,k} + \Phi_{nD}^1(\mu_{nD}^1)N_{nD,k}+\Phi_{R}^1(\mu_{R}^1)N_{R,k}}{N_k}\right)\nonumber \\
 &&\cdot \left(\frac{\Phi_{mD}^2(\mu_{mD}^2)N_{mD,k} + \Phi_{nD}^2(\mu_{nD}^2)N_{nD,k}+\Phi_{R}^2(\mu_{R}^2)(N_k-N_{mD,k}-N_{nD,k})}{N_k} \right) \nonumber \\
 &&+ \left(1-\frac{\Phi_{mD}^1(\mu_{mD}^1)N_{mD,k} + \Phi_{nD}^1(\mu_{nD}^1)N_{nD,k}+\Phi_{R}^1(\mu_{R}^1)N_{R,k}}{N_k}\right)\nonumber \\&&\cdot\left(\frac{\Phi_{mD}^3(\mu_{mD}^3)N_{mD,k} + \Phi_{nD}^3(\mu_{nD}^3)N_{nD,k}+\Phi_{R}^3(\mu_{R}^3)(N_k-N_{mD,k}-N_{nD,k})}{N_k} \right)\beta .
\end{eqnarray}

The second derivative with respect to the number of minority voters in a district is 
\begin{equation}\label{app_eq-exp-min-utility-2nd-open}
\frac{\partial^2 E[\mu_{mD}\mid k]}{\partial N_{mD,k}^2}= \frac{2\overbrace{(\Phi_{mD}^1(\mu_{mD}^1)-\Phi_{R}^1(\mu_{R}^1))}^{(+)}\delta}{N_k^2},
\end{equation}
where
\begin{equation}\label{appendix_delta}
\delta\equiv \Phi_{mD}^2(\mu_{mD}^2)-\Phi_{R}^2(\mu_{R}^2) -\beta(\Phi_{mD}^3(\mu_{mD}^3)-\Phi_{R}^3(\mu_{R}^3)).
\end{equation}
Evaluating the convexity/concavity, we write the condition in terms of $\beta$:
\begin{equation}
\frac{\partial^2 E[\mu_{mD}\mid k]}{\partial N_{mD,k}^2} > 0 \Leftrightarrow \beta < \frac{ \Phi_{mD}^2(\mu_{mD}^2)-\Phi_{R}^2(\mu_{R}^2)}{\Phi_{mD}^3(\mu_{mD}^3)-\Phi_{R}^3(\mu_{R}^3)}.
\end{equation}

\subsection{Proof of Proposition \ref{proposition_ideological-districting-distribution}}\label{proof_proposition_ideological-districting-distribution}

Let $s_k\equiv N_{mD,k}/N_k$ with $N_k$ identical across all districts and $\sum_k N_k s_k=N_{mD}$ be fixed. Under the selection channel with fixed matchups, minority ideological welfare can be rewritten as
\begin{equation}
W^{s}(D)= \sum_{k=1}^Ks_kN_k\cdot E[\mu_{mD}|k] = \sum_{k=1}^K N_k\cdot w^s(s_k), \qquad
w^s(s_k)\equiv s_k\,E[\mu_{mD}\mid s_k].
\end{equation}
Lemma~\ref{lemma-distributive-districting-payoff-simplex} implies that $w^s(s_k)$ is convex when (i) $\beta<\beta^C$ and $\Delta_k^C>0$, (ii) $\beta>\beta^C$ and $\Delta_k^C<0$, (iii) $\beta<\beta^O$ and concave when the opposite. 

\paragraph{1) Convexity}

Let $s_{k_1}>s_{k_2}$ and consider a mean-preserving transfer $\delta>0$ from district $k_2$ to $k_1$ (for $\delta$ small enough).

If $w^s(s_k)$ is convex, then $w^s(s_{k_1}+\delta) + w^s(s_{k_2}-\delta) \geq w^s(s_{k_1}) + w^s(s_{k_2})$, so concentrating minority voters weakly increases $W^s(D)$. By repeating the process, minority welfare is maximized at boundary allocations -- i.e., by concentrating minority voters as much as feasible.

\paragraph{2) Concavity}

Let $s_{k_1}>s_{k_2}$ and consider a mean-preserving transfer $\delta>0$ from district $k_1$ to $k_2$ (again for small enough $\delta$).

If $w^s(s_k)$ is concave, then $w^s(s_{k_1}-\delta) + w^s(s_{k_2}+\delta) \geq w^s(s_{k_1}) + w^s(s_{k_2})$, so dispersing minority voters weakly increases $W^s(D)$. By repeating the process, minority welfare is maximized under more equal allocations, i.e., by dispersing minority voters as evenly as possible.

\subsection{Decomposition of Minority Welfare Curvature}\label{appendix_minority-welfare-curvature-details}

Minority welfare follows from 
\begin{eqnarray}
W(D) &=& \underbrace{\sum_{k=1}^KN_{mD,k}\cdot \sum_{j\in \Theta}\Psi_j(k)\cdot\mu_{mD,j}}_{W^s(D)} + \underbrace{\sum_{k=1}^KN_{mD,k}\cdot \sum_{j\in \Theta}\Psi_j(k)\cdot\kappa_{mD}\cdot\frac{b_{mD,j,k}^{1-\epsilon}}{1-\epsilon}}_{W^c(D)}\\
&=& \sum_{k=1}^KN_ks_k\cdot \sum_{j\in \Theta}\Psi_j(k)\cdot\mu_{mD,j} + \sum_{k=1}^KN_ks_k\cdot \sum_{j\in \Theta}\Psi_j(k)\cdot\kappa_{mD}\cdot\frac{b_{mD,j,k}^{1-\epsilon}}{1-\epsilon},
\end{eqnarray}
where $s_k\equiv N_{mD,k}/N_k$ and $N_k$ is fixed. 

When matchups are allowed to adjust endogenously, $\Psi_j(k)$, $\mu_i^e$, and therefore $b_{mD,j,k}$ become functions of $s_k$ through the equilibrium chain
\begin{equation}
s_k \rightarrow \Psi_j(k) \rightarrow \mu_i^e \rightarrow \Phi^e_i(\mu_i^e) \text{ and } \phi_i(\mu_i^e) \rightarrow \pi_i^e \rightarrow b_{mD,j,k}.
\end{equation}

\paragraph{Curvature Decomposition}
Differentiating twice with respect to $s_k$ yields
\begin{equation}\label{eq:decomposition-final-second}
\frac{\partial^2 W(D)}{\partial s_k^2} = \frac{\partial^2 W^s(D)}{\partial s_k^2} + \left.\frac{\partial^2 W^c(D)}{\partial s_k^2}\right|_{\substack{\text{fixed}\\\text{matchup}}} + I_k,
\end{equation}
where the second term corresponds to Section~\ref{section_substantive-representation}, which holds matchups fixed, and
\begin{equation}
I_k \equiv \frac{\partial^2 W^c(D)}{\partial s_k^2} - \left.\frac{\partial^2 W^c(D)}{\partial s_k^2}\right|_{\substack{\text{fixed}\\\text{matchup}}} 
\end{equation}
collects the additional terms arising from endogenous feedback: the additional curvature arising from the dependence of 
$\Psi_j(k)$, $\mu_i^e$, and therefore $\pi_i^e$ and $b_{mD,j,k}$, on $s_k$.

The curvature of the selection channel follows from Section~\ref{section_descriptive-representation},
\begin{equation}\label{eq:decomposition-selection-second}
\frac{\partial^2 W^s(D)}{\partial s_k^2} = N_k\left( 2\sum_{j\in\Theta}\mu_{mD,j}\frac{\partial \Psi_j(k)}{\partial s_k}
+ s_k\sum_{j\in\Theta}\mu_{mD,j}\frac{\partial^2 \Psi_j(k)}{\partial s_k^2}
\right).
\end{equation}

The curvature of the competition channel under fixed matchups relates to Section~\ref{section_substantive-representation},
holding the general-election matchup fixed (treating $\mu_i^e=\mu_i^G$ as constant),
\begin{eqnarray}\label{eq:decomposition-competition-second}
\left.\frac{\partial^2 W^c(D)}{\partial s_k^2}\right|_{\substack{\text{fixed}\\\text{matchup}}}
&=& N_k \kappa_{mD}\sum_{j\in\Theta} \bar{\Psi}_j
\Bigg[
2\, b_{mD,j,k}^{-\epsilon}
\left.\frac{\partial b_{mD,j,k}}{\partial s_k}\right|_{\substack{\text{fixed}\\\text{matchup}}}
+ s_k\Bigg(
-\epsilon\, b_{mD,j,k}^{-\epsilon-1}
\left(
\left.\frac{\partial b_{mD,j,k}}{\partial s_k}\right|_{\substack{\text{fixed}\\\text{matchup}}}
\right)^2
\nonumber\\
&& + b_{mD,j,k}^{-\epsilon}
\left.\frac{\partial^2 b_{mD,j,k}}{\partial s_k^2}\right|_{\substack{\text{fixed}\\\text{matchup}}}
\Bigg)
\Bigg].
\end{eqnarray}
Although equilibrium platforms coincide across candidates within a given general-election environment (Lemma~\ref{lemma-platforms}), the induced matchup $\mu_i^e$ affects group power $\pi_i^e$ and benefits $b_{i,j,k}$, and hence, platforms may differ across potential electoral outcomes. 

The interaction effect, by definition, $I_k$ captures only the additional terms that arise because $\Psi_j(k)$ and $\mu_i^e$ vary with $s_k$:
\begin{eqnarray}\label{eq:interaction-detailed-correct}
I_k &=& N_k \kappa_{mD} \Bigg\{
2\sum_{j\in\Theta}\frac{\partial \Psi_j(k)}{\partial s_k}\cdot\frac{b_{mD,j,k}^{1-\epsilon}}{1-\epsilon}
+ 2\sum_{j\in\Theta}\Psi_j(k)\cdot b_{mD,j,k}^{-\epsilon}
\left(\frac{\partial b_{mD,j,k}}{\partial s_k}
-\left.\frac{\partial b_{mD,j,k}}{\partial s_k}\right|_{\substack{\text{fixed}\\\text{matchup}}}\right)
\nonumber\\
&&+ s_k\sum_{j\in\Theta}
\Bigg[
\frac{\partial^2 \Psi_j(k)}{\partial s_k^2}\cdot\frac{b_{mD,j,k}^{1-\epsilon}}{1-\epsilon}
+ 2\frac{\partial \Psi_j(k)}{\partial s_k}\cdot b_{mD,j,k}^{-\epsilon}\cdot\frac{\partial b_{mD,j,k}}{\partial s_k}
\nonumber\\
&&
+ \Psi_j(k)\Bigg(
-\epsilon\,b_{mD,j,k}^{-\epsilon-1}
\left[\left(\frac{\partial b_{mD,j,k}}{\partial s_k}\right)^2
-\left(\left.\frac{\partial b_{mD,j,k}}{\partial s_k}\right|_{\substack{\text{fixed}\\\text{matchup}}}\right)^2\right]
\nonumber\\
&&
+ b_{mD,j,k}^{-\epsilon}
\left[\frac{\partial^2 b_{mD,j,k}}{\partial s_k^2}
-\left.\frac{\partial^2 b_{mD,j,k}}{\partial s_k^2}\right|_{\substack{\text{fixed}\\\text{matchup}}}\right]
\Bigg)
\Bigg]
\Bigg\}.
\end{eqnarray}

Using the chain rule,
\begin{equation}\label{eq:interaction-targeting}
\frac{\partial b_{mD,j,k}}{\partial s_k} = \sum_{i\in\Theta} \frac{\partial b_{mD,j,k}}{\partial \pi_i^e}
\frac{\partial \pi_i^e}{\partial s_k} + \sum_{i\in\Theta} \frac{\partial b_{mD,j,k}}{\partial \mu_i^e}
\frac{\partial \mu_i^e}{\partial s_k}.
\end{equation}
Because group power satisfies
\begin{equation}\label{eq:interaction-power}
\frac{\partial \pi_i^e}{\partial s_k} = \frac{1}{\epsilon} \pi_i^e \frac{\phi_i'(\mu_i^e)}{\phi_i(\mu_i^e)}
\frac{\partial \mu_i^e}{\partial s_k}.
\end{equation}
The interaction term depends directly on
\begin{equation}\label{eq:interaction-three-terms}
\frac{\partial \Psi_j(k)}{\partial s_k}, \text{ } \frac{\partial \mu_i^e}{\partial s_k}, \text{ and } \frac{\phi_i'(\mu_i^e)}{\phi_i(\mu_i^e)},
\end{equation}
which determine how district composition translates into changes in equilibrium pivotality and targeting incentives.

\paragraph{Curvature of Interaction Effect}

The sign of $I_k$, and therefore the interaction contribution to the curvature of $W(D)$, is determined by how district composition jointly affects electoral probabilities and minority targeting incentives. From equation~\eqref{eq:interaction-detailed-correct} and the chain-rule relations in \eqref{eq:interaction-targeting}–\eqref{eq:interaction-power}, two mechanisms govern its sign $I_k$. Changes in minority concentration alter
\begin{itemize}
\item Electoral outcomes through $\partial \Psi_j(k)/\partial s_k$ and $\partial^2 \Psi_j(k)/\partial s_k^2$. If increasing $s_k$ shifts probability mass toward candidates $j$ under which minority distributive utility $\frac{b_{mD,j,k}^{1-\epsilon}}{1-\epsilon}$ is larger, this probability reweighting contributes positively to $I_k$ (reinforcing convexity of $W(D)$). Conversely, if composition shifts increase the likelihood of outcomes with lower minority transfers, this component contributes negatively (reinforcing concavity of $W(D)$).

\item Endogenous targeting effects operate through matchup sensitivity $\partial \mu_i^e/\partial s_k$ and responsiveness elasticity $\phi_i'(\mu_i^e)/\phi_i(\mu_i^e)$. Because group power satisfies $\partial \pi_i^e/\partial s_k =
\frac{1}{\epsilon}\pi_i^e\frac{\phi_i'(\mu_i^e)}{\phi_i(\mu_i^e)}\frac{\partial \mu_i^e}{\partial s_k}$, interaction curvature is positive (reinforcing convexity of $W(D)$) whenever the induced change in matchups raises minority group power and transfers are increasing in group power (i.e., $\partial b_{mD,j,k}/\partial \pi_{mD}^e>0$). It is negative (reinforcing concavity of $W(D)$) when compositional changes reduce minority pivotality or shift matchups toward environments with weaker targeting incentives.
\end{itemize}
Taken together, $I_k>0$ (reinforcing feedback) when increasing minority concentration both (i) increases the probability of minority-favorable electoral outcomes and (ii) strengthens minority pivotality in distributive competition. By contrast, $I_k<0$ (offsetting feedback) when concentration either shifts probability mass toward minority-unfavorable outcomes or
reduces marginal responsiveness and group power. 

These conditions are sufficient but not necessary; in general, the sign depends on the relative magnitude of the terms in \eqref{eq:interaction-detailed-correct}.

\paragraph{Electoral Stability versus Tipping-Point Sensitivity} 

The distinction between safe and tipping districts corresponds to whether the equilibrium objects entering the curvature decomposition in equation~\eqref{eq:decomposition-final-second} respond weakly or strongly to changes in $s_k$. In particular:
\begin{itemize}
\item[(i)] Condition (i) in Definition~\ref{definition-safe-districts} concerns the selection derivatives $\partial \Psi_j(k)/\partial s_k$ and $\partial^2 \Psi_j(k)/\partial s_k^2$, which enter directly into the selection curvature \eqref{eq:decomposition-selection-second} and the interaction term \eqref{eq:interaction-detailed-correct}. When these derivatives are small, compositional changes have limited effects on who runs or wins.

\item[(ii)] Condition (ii) concerns the sensitivity of matchups $\mu_i^e$ to district composition. The derivatives $\partial \mu_i^e/\partial s_k$ and $\partial^2 \mu_i^e/\partial s_k^2$ enter the interaction term through the chain rule for targeting in \eqref{eq:interaction-targeting}. When $\partial \mu_i^e/\partial s_k$ is small, ideological distances shift only modestly across districts.

\item[(iii)] Condition (iii) concerns voter responsiveness. Because group power satisfies equation~\eqref{eq:interaction-power}, the responsiveness elasticity $\phi_i'(\mu_i^e)/\phi_i(\mu_i^e)$ amplifies the effect of matchup changes on $\pi_i^e$. Hence the objects summarized in \eqref{eq:interaction-three-terms} determine how composition affects equilibrium pivotality and targeting incentives.

\end{itemize}

Together, these stability conditions imply that the interaction term $I_k$ in equation~\eqref{eq:decomposition-final-second} remains small when selection probabilities, ideological matchups, and marginal responsiveness respond weakly to changes in $s_k$. This follows directly from the structure of $I_k$ in \eqref{eq:interaction-detailed-correct} and the chain-rule relationships in \eqref{eq:interaction-targeting}–\eqref{eq:interaction-power}. By contrast, in tipping districts—such as those near nomination or general-election cutoffs—one or more of the derivatives in \eqref{eq:interaction-three-terms} is large, so the interaction term becomes first-order and may dominate the direct curvature components in \eqref{eq:decomposition-final-second}.

\subsection{Proof of Lemma \ref{lemma_general_equilibrium}}\label{proof_lemma_general_equilibrium}

From the curvature decomposition in equation~\eqref{eq:decomposition-final-second}, total curvature satisfies
\begin{equation}
W_{ss}(D) = C_k + I_k, \qquad C_k \equiv W_{ss}^s + W_{ss}^{c,0}.
\end{equation}

We have three cases to consider.

\paragraph{1) Fixed-Matchup Benchmark} If matchups are held fixed, then by definition $I_k=0$, so $W_{ss}(D)=C_k$. Alignment arises when $\sign(W_{ss}^s)=\sign(W_{ss}^{c,0})$, and divergence arises when $\sign(W_{ss}^s)\neq\sign(W_{ss}^{c,0})$. This follows directly from the additive separation of welfare into selection and competition components characterized in Propositions~\ref{proposition_distributive-districting-distribution} and~\ref{proposition_ideological-districting-distribution}.

\paragraph{2) Safe Districts} By Definition~\ref{definition-safe-districts}, the derivatives governing equilibrium feedback,
$\partial \Psi_j(k)/\partial s_k$, $\partial \mu_i^e/\partial s_k$, and $\phi_i'(\mu_i^e)/\phi_i(\mu_i^e)$, are uniformly small in a neighborhood of $s_k$. Appendix~\ref{appendix_minority-welfare-curvature-details} shows that the interaction term $I_k$ consists of cross-partials proportional to these derivatives. Hence $|I_k|$ is second-order relative to the direct curvature benchmark $C_k$.

If $|C_k|$ is bounded away from zero, then for sufficiently small feedback $|I_k|<|C_k|$, implying $\sign(C_k+I_k)=\sign(C_k)$.
Thus, alignment or divergence under fixed matchups is locally preserved.

\paragraph{3) Tipping Districts} By Definition~\ref{definition-tipping-districts}, at least one of the feedback derivatives
is large in magnitude. From the structure of $I_k$ in \eqref{eq:interaction-detailed-correct} and the chain-rule relations in
\eqref{eq:interaction-targeting}–\eqref{eq:interaction-power}, $I_k$ may therefore be first-order relative to $C_k$.

If $\sign(I_k)=\sign(C_k)$, then $\sign(C_k+I_k)=\sign(C_k)$ and feedback reinforces the fixed-matchup benchmark.

If $\sign(I_k)\neq\sign(C_k)$ and $|I_k|>|C_k|$, then $\sign(C_k+I_k)\neq\sign(C_k)$, so feedback overturns the benchmark.

\subsection{Proof of Lemma \ref{lemma_tipping_region}}\label{proof_lemma_tipping_region}

By Lemma~\ref{lemma_general_equilibrium}, the interaction term $I_k$ is generated only by endogenous changes in (i) electoral outcome probabilities $\Psi_j(k)$ and (ii) the induced general-election environment $\mu^e$ (hence pivotality and group power) along the equilibrium chain in \eqref{eq:detailed-mechanism}. Under the standard single-index probabilistic voting structure with a unique cutoff, write each outcome probability as $\Psi_j(k)=F_j(\Delta_j(s_k))$, where $\Delta_j(s_k)$ is a monotone scalar index and $F_j(.)$ is smooth. Then
\begin{equation}
\frac{\partial \Psi_j(k)}{\partial s_k}=F_j'(\Delta_j(s_k))\,\Delta_j'(s_k),
\end{equation}
so $\partial \Psi_j(k)/\partial s_k$ is quantitatively relevant only when $\Delta_j(s_k)$ lies near its unique cutoff (where the pivot density $F_j'$ is large), and is negligible when $s_k$ is far from that cutoff. In particular, at low minority shares $s_k$ the general-election environment is effectively pinned to the $nD-R$-matchup, while at high $s_k$ it is pinned to the $mD-R$-matchup. In both regions, marginal changes in $s_k$ do not move the equilibrium matchup, implying $\partial \mu_i^e/\partial s_k\approx 0$ and therefore $\partial \pi_i^e/\partial s_k\approx 0$ via $\partial \pi_i^e/\partial s_k=\frac{\pi_i^e}{\epsilon}\frac{\phi_i'(\mu_i^e)}{\phi_i(\mu_i^e)}\frac{\partial \mu_i^e}{\partial s_k}$. By contrast, at intermediate $s_k$ near the cutoff, small compositional changes shift winning probabilities and hence can switch the relevant general-election matchup, making $\partial \Psi_j(k)/\partial s_k$ and $\partial \mu_i^e/\partial s_k$ first-order. Finally, because endogenous objects enter $I_k$ only through the two components in \eqref{eq:Ik-schematic}, probability reweighting terms proportional to $\partial \Psi_j(k)/\partial s_k$ and pivotality/targeting terms proportional to $\partial \pi_i^e/\partial s_k$. It follows that $I_k$ is first-order only on the connected interval $(\underline{s},\overline{s})$ where the cutoff is locally significant, and is second-order elsewhere.

\subsection{Proof of Proposition \ref{proposition-total-welfare-voter-distribution}}\label{proof_proposition-total-welfare-voter-distribution}

Let $s_k\equiv N_{mD,k}/N_k$ with $N_k$ identical across all districts and $\sum_k N_k s_k=N_{mD}$ be fixed. Define district-level minority payoffs
\begin{equation}
w^s(s_k)\equiv s_k\sum_{j\in\Theta}\Psi_j(s_k)\mu_{mD,j} \quad\text{and}\quad
w^{c,0}(s_k)\equiv s_k\sum_{j\in\Theta}\bar{\Psi}_j\kappa_{mD}\frac{b^{0}_{mD,j,k}(s_k)^{1-\epsilon}}{1-\epsilon}.
\end{equation}
Let $w^s(s_k)+w^{c,0}(s_k)$ denote the fixed-matchup benchmark payoff and define the interaction payoff $w^I$ by $w_{ss}^I(s_k)=I_k(s_k)$. Then total minority welfare can be written as
\begin{equation}
W(D)=\sum_{k=1}^K N_k\,w(s_k),\qquad w(s_k)\equiv w^s(s_k) + w^{c,0}(s_k) + w^I(s_k),
\end{equation}
so that $w_{ss}(s_k)=C_k(s_k)+I_k(s_k)$.

\paragraph{1) Global Convexity}

Suppose $w_{ss}(s_k) > 0$ for all feasible $s_k$. Then $w(.)$ is convex, and hence $W(D)$ is convex on the feasible set. Let $s_{k_1}>s_{k_2}$ and consider a mean-preserving transfer $\delta>0$ from district $k_2$ to $k_1$ (for $\delta$ small enough).

By convexity, $w(s_{k_1}+\delta) + w(s_{k_2}-\delta) \geq w(s_{k_1}) + w(s_{k_2})$, so concentrating minority voters weakly increases $W(D)$. By repeating the process, minority welfare is maximized at more unequal feasible allocations -- i.e., by concentrating minority voters as much as feasible.

\paragraph{2) Global Concavity}

Suppose $w_{ss}(s_k) < 0$ for all feasible $s_k$. Then $w(.)$ is concave and hence $W(D)$ is concave on the feasible set. Let $s_{k_1}>s_{k_2}$ and consider a mean-preserving transfer $\delta>0$ from district $k_1$ to $k_2$ (again for small enough $\delta$).

By concavity, $w(s_{k_1}-\delta) + w(s_{k_2}+\delta) \geq w(s_{k_1}) + w(s_{k_2})$, so dispersing minority voters weakly increases $W(D)$. By repeating the process, minority welfare is maximized at more equal feasible allocations, i.e., by dispersing minority voters as evenly as possible.

\paragraph{3) Curvature Changes}

Suppose $w_{ss}(s_k)$ changes sign over feasible $s_k$, then $w(.)$ is neither globally convex nor globally concave on the feasible set.  Global convexity (concavity) obtains whenever $w_{ss}(s_k)=C_k(s_k)+I_k(s_k)>0$ ($<0$ respectively) for all feasible $s_k$. If instead $w_{ss}(s_k)=C_k(s_k)+I_k(s_k)>0$ for some feasible $s_k$ but $w_{ss}(s_k)=C_k(s_k)+I_k(s_k)<0$ for others, then by continuity there exists at least one curvature reversal. Such reversals can be driven either by 
\begin{enumerate}[label=(\alph*)]
\item benchmark nonmonotonicity, $C_k(s_k)=W^s_{ss}(s_k)+W^{c,0}_{ss}$ changing sign along $s_k$ due to divergence between the selection and competition curvatures (Table~\ref{tbl:summary-channels-voter-distribution-detailed}); or

\item feedback overturning in tipping districts, where $I_k$ is first-order (Lemma~\ref{lemma_tipping_region}) and $\sign(I_k)\neq\sign(C_k)$ with $I_k$ sufficiently large to flip -- i.e., $\sign(C_k)\neq\sign(C_k+I_k)$.
\end{enumerate}

\newpage
\section{Appendix: Illustrations and Simulations}\label{section_appendix_illustrations}

\subsection{Sample State and Five Districts}\label{section_appendix_numerics_state}

The example of Figure~\ref{fig:SampleState} is created with $S=(.36, 0.26, .38)$ and a valid districting matrix for five districts can be illustrated by 
\begin{eqnarray*}
    \left(
    \begin{matrix}
        0.19   & 0.6   & 0.21   \\
        0.33   & 0.05   & 0.62    \\
        0.45 & 0.1 & 0.45 \\
        0.14   & 0.43   & 0.42 \\
        0.65 & 0.13 & 0.22
    \end{matrix}
    \right).
\end{eqnarray*}

\end{document}